\documentclass[conference]{IEEEtran}
\pdfoutput=1
\IEEEoverridecommandlockouts
\usepackage[table,dvipsnames]{xcolor} % Move it up
\usepackage{cite}
\usepackage{amsmath,amssymb,amsfonts}
\usepackage{algorithmic}
\usepackage{graphicx}
\usepackage{textcomp}

\usepackage{algorithm}
\usepackage{multirow}
\usepackage{array}

\usepackage{subcaption}
\usepackage{graphicx}
\usepackage{textcomp}
\usepackage{booktabs}

\def\BibTeX{{\rm B\kern-.05em{\sc i\kern-.025em b}\kern-.08em
    T\kern-.1667em\lower.7ex\hbox{E}\kern-.125emX}}
\begin{document}

\title{\huge CP-AgentNet: Autonomous and Explainable Communication Protocol Design Using Generative Agents
}

\author{\IEEEauthorblockN{Dae Cheol Kwon, Xinyu Zhang}
\IEEEauthorblockA{
\textit{University of California San Diego}\\
\{dckwon, xyzhang\}@ucsd.edu}
}
\maketitle

\IEEEtitleabstractindextext{%
\begin{abstract}
Although DRL (deep reinforcement learning) has emerged as a powerful tool for making better decisions than existing hand-crafted communication protocols, it faces significant limitations: 1) Selecting the appropriate neural network architecture and setting hyperparameters are crucial for achieving desired performance levels, requiring domain expertise. 2) The decision-making process in DRL models is often opaque, commonly described as a 'black box.' 3) DRL models are data hungry. In response, we propose CP-AgentNet, the first framework designed to use generative agents for developing communication network protocols. This approach addresses these challenges by creating an autonomous system for protocol design, significantly reducing human effort. We developed LLMA (LLM-agents-based multiple access) and CPTCP (CP-Agent-based TCP) for heterogeneous environments. Our comprehensive simulations have demonstrated the efficient coexistence of LLMA and CPTCP with nodes using different types of protocols, as well as enhanced explainability.
\end{abstract}
}

\IEEEdisplaynontitleabstractindextext  % This ensures the abstract is displayed.

\section{Introduction}\label{Introduction}
\begin{figure}
  \centering
  \centerline{\includegraphics[width=1\linewidth]{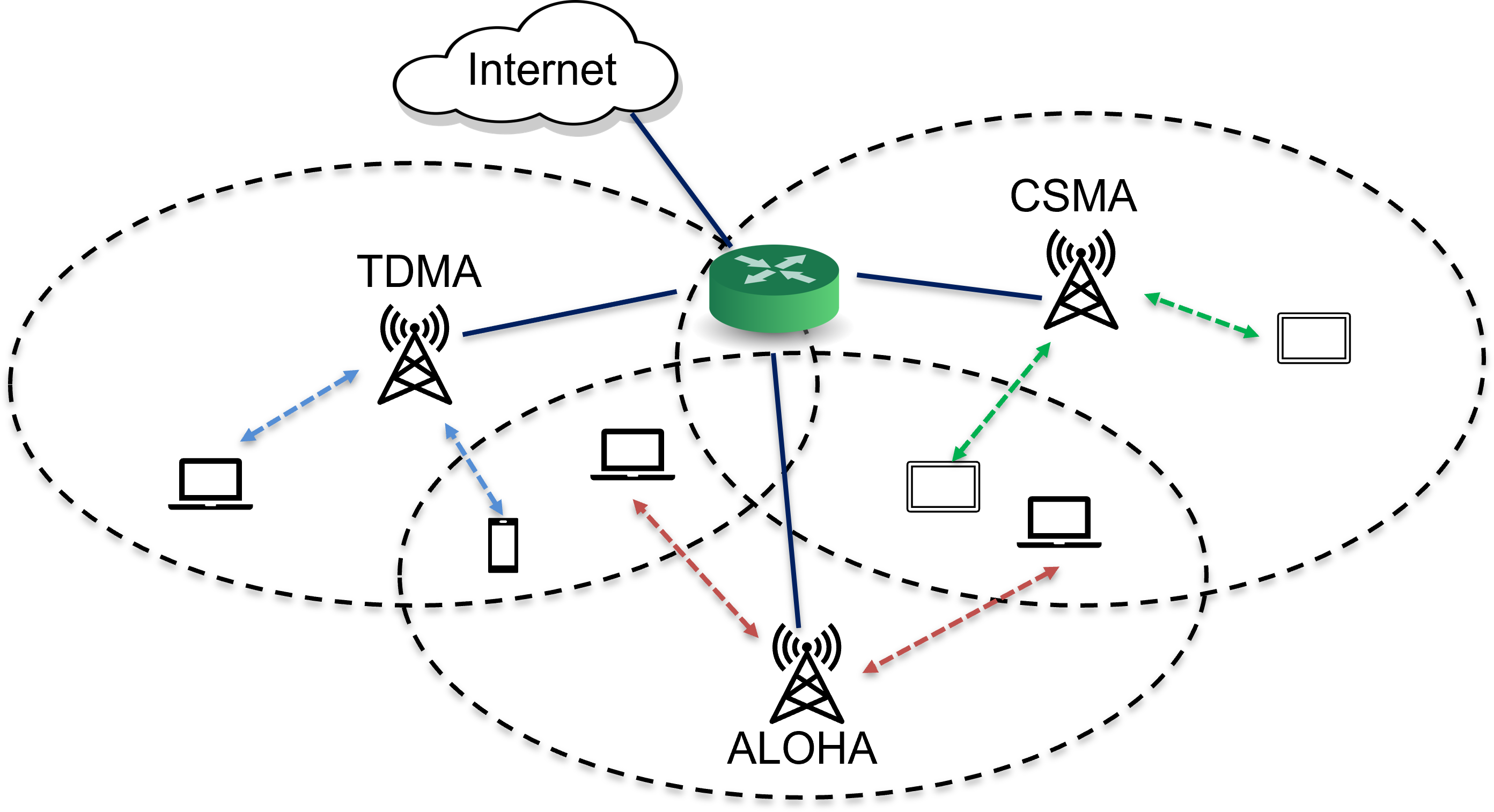}}
  \caption{Autonomous and explainable protocol design with stability is achieved via CP-AgentNet}
  \label{fig:heteroNet}
\end{figure}

Conventional communication protocols are meticulously established through standardization processes. However, challenges arise from the explosive demands for data traffic and ever-increasing network heterogeneity, as illustrated in Fig.~\ref{fig:heteroNet}.  Typically designed for a single type of network, standard communication protocols lack the adaptability to coexist in diverse networks. Enabling operations across heterogeneous networks concurrently necessitates the creation of a new protocol for each coexistence scenario, demanding significant expertise and effort.

DRL (deep reinforcement learning) has emerged as a powerful tool for devising communication protocols, which are essentially decision-making algorithms executed by the agents, i.e., network nodes. Despite their potential, DRL methods have several drawbacks. First, choosing an appropriate neural network architecture and tuning hyperparameters are crucial yet often require tedious trial-and-error experimentation. Additionally, neural network architectures optimized for specific training environments often lose effectiveness when the environment changes, potentially requiring costly re-design or retraining, thus further limiting their practical adaptability.  Second, the decision-making process in DRL models is is often considered a ``black box'' \cite{blackbox}. In addition, DRL models are data hungry. It entails not only labor-intensive data collection, but also labeling and preprocessing. Data acquisition may induce non-trivial overhead, disturbing the normal network operations. 

In this paper, we introduce CP-AgentNet, a framework designed to create communication protocols using generative agents, specifically those empowered by LLMs, hereafter referred to as LLM-agents. CP-AgentNet addresses the inherent limitations of both handcrafted and DRL-based protocols. By leveraging the in-context learning capabilities of LLMs, these LLM-agents can quickly interpret and adapt to instructions provided through context or interaction sequences, eliminating the need for explicit retraining when conditions change. Our framework requires crafting only a few demonstrations, capitalizing on the strong generalization abilities of LLM-agents to efficiently guide their training. This significantly reduces the effort involved in selecting neural network architectures, tuning hyperparameters, and gathering, labeling, and preprocessing large datasets. Additionally, since the outputs generated by LLM-agents are in natural language, they can be easily interpreted and understood by humans, greatly enhancing interpretability and facilitating straightforward interactions.

However, despite substantial advantages of using LLM-agents, their use still presents significant challenges. First and foremost, although the LLM-agent is capable of handling tasks across various disciplines, a single LLM-agent struggles to manage multiple tasks simultaneously \cite{SelfOrganizedAgents}. Despite its capability in a broad range of topics, it often lacks the deep, specialized knowledge required for complex tasks. In particular, an agent in this framework must make transmission decisions while concurrently monitoring the network environment, adding complexity to its operational tasks. To address this, we employ multi-agent role-play, an approach already proven effective for tasks in programming and mathematical reasoning. Harnessing multiple agents proves advantageous as each agent plays a distinct role, simplifying complex tasks\cite{self-collaboration}. Additionally, this approach significantly enhances explainability by making the decision-making process traceable.

Second, establishing an autonomous design framework using LLM-agents presents unique challenges, primarily due to the absence of explicit optimization methods such as gradient descent, which are commonly used in traditional DRL to systematically improve policies. Existing LLM-agent frameworks often compensate for this limitation by relying heavily on explicit human feedback \cite{autogen, webgpt}, thereby requiring substantial user effort. To address this fundamental challenge, we introduce \textit{Progressive Strategy Augmentation (PSA)}, an approach specifically designed for LLM-agents. Rather than updating parameters via gradient-based optimization, PSA utilizes a structured self-reflection mechanism \cite{reflexion} to incrementally refine strategies. Guided by carefully crafted prompts, the LLM-agent autonomously reviews previous episodes, identifies underlying reasons for suboptimal outcomes, and progressively revises its strategies. The refined strategies obtained through multiple self-reflection episodes are subsequently aggregated, and redundant or conflicting components are systematically identified and removed. In essence, PSA serves as a symbolic counterpart to gradient descent, leveraging the generative reasoning and in-context learning capabilities inherent to LLMs. Another method we employ for autonomous design is \textit{Autonomous Strategy Implementation (ASI)}. There are instances where responses from the LLM may be unpredictable, especially when the prompt cannot specify the response format. In such cases, ASI is crucial for maintaining autonomy throughout the design process. By utilizing the programming capabilities of the LLM, the programming assistant can autonomously implement responses.

Last but not least, consistent outputs are not guaranteed with LLMs due to their inherent nature of predicting the probability distribution over the next token. This variability can lead to incorrect planning, or the programming scripts generated by the programming assistant might sometimes result in compile errors due to a misunderstanding of the requirements. To enhance consistency, we employ an LLM ranker\cite{ranker} and a compilation checker. Inspired by \textit{The Wisdom of Crowds}\cite{wisdom}, which posits that ``large groups of people are collectively smarter than individual experts when it comes to problem-solving, decision making, innovating and predicting'', we propose the use of an LLM ranker. The LLM ranker processes multiple output candidates along with their corresponding prompts and selects the best candidate as the final output. The compilation checker verifies input parameters and syntax errors before execution; if it detects errors in the programming function, it triggers a new query to the LLM to generate the script again. These methods significantly improve the stability of CP-AgentNet.

%Overall, CP-AgentNet enables autonomous and explainable protocol design with stability: decision-making by multiple agents enhances explainability; self-examination and self-implementation processes ensure autonomy; and stability is maintained by the LLM ranker and compilation checker. These elements are summarized in Fig.~\ref{fig:DRL_LLM_agent}. 

As a use case of CP-AgentNet, we designed the LLMA (LLM-agents-based multiple access) protocol for heterogeneous networks. To verify the effectiveness of the LLMA protocol, we conducted comprehensive simulations across various scenarios. Specifically, we first established the ideal operation of an AWARE node—a hypothetical node equipped with complete knowledge of the environment. We then compared the performance of the LLMA node to that of the AWARE node by measuring RMSE (root mean square error). As a benchmark, we also assessd the performance of the DLMA node. Our results indicate that the LLMA node performs closer to the ideal than the DLMA node.
Remarkably, in dynamic environments where nodes using different protocols intermittently join or leave, the LLMA node adapts more swiftly and exhibits a significantly lower RMSE value of 0.0476--just 21.0\% of the RMSE of 0.227 resulting from the DLMA node.

In addition, we designed CPTCP (CP-Agent-based Transmission Control Protocol) to coexist with different types of TCP algorithms. There are mainly two types of TCP algorithms: loss-based and delay-based TCP. We show that CPTCP enables effective coexistence with both types of TCP algorithms. Moreover, this demonstrates that our framework is not limited to designing a specific protocol.

In summary, the main contributions of this paper are:
\begin{itemize}
\item We introduce CP-AgentNet, a framework designing communication protocols using generative agents. CP-AgentNet employs multi-agent role-play to efficiently process tasks and facilitate the explainable design of communication protocols. 

\item We propose PSA and ASI to facilitate autonomous design. These enable LLM-agents to develop strategies and implement them to optimize network performance, without the need for human intervention.

\item We have developed LLMA and CPTCP, specifically designed for a heterogeneous environment using CP-AgentNet. We demonstrate that nodes using these protocols can efficiently coexist with other nodes operating different protocols.
\end{itemize}
\section{Background}\label{Background}

The recent advancements in LLMs have enhanced their application in agents—computer systems designed to perform autonomous actions in specific environments to achieve predefined objectives \cite{autoagent}. By leveraging the capabilities of LLMs, these agents can make more informed decisions and effectively execute tasks; we refer to these as LLM-empowered agents, or LLM-agents. 
The agent queries the LLM to handle the given task. Based on the LLM's response, the agent makes a decision to act within the environment. The environment provides feedback or a reward to the agent, which is subsequently utilized to inform the LLM in planning the next action. In this process, agents can utilize external tools such as memory to enhance their functionality, particularly in tasks requiring context awareness. Integrating memory with LLMs enables these agents to retain information over extended periods, thereby improving the continuity of interactions and allowing for more informed decisions based on past experiences.
LLM-agents have demonstrated effectiveness across a diverse array of tasks, including reasoning, embodied tasks, and code generation. Additionally, they are utilized in various disciplines such as computer science\cite{metagpt,self-collaboration,qin2023toolllm,safdari2023personality}, robotics\cite{wang2023voyager,ghost,robotic1}, political science\cite{social}, economics\cite{economy}, social simulation\cite{generative_agent}, jurisprudence\cite{law1,law2}, and so on. However, in the field of communication networks, the use of LLMs is primarily confined to specific tasks like classification and Q\&A\cite{Bert-QA, LLM_standard, LLM_telecom, LLM_small, teleqna, TKG}.
\begin{figure*}
  \centering
  \includegraphics[width=\textwidth]{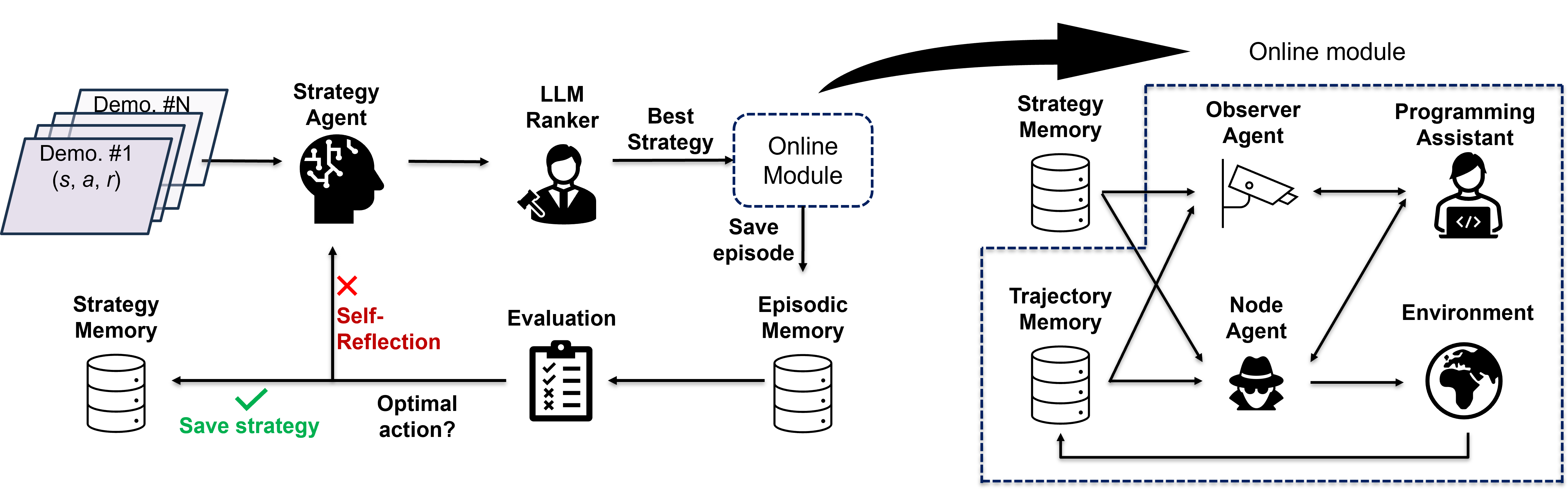}
  \caption{The workflow of CP-AgentNet. Left: Offline stage, Right: Online stage.}
  \label{fig:CP-AgentNet}
\end{figure*}

\section{System Design}\label{CP-AgentNet}
\subsection{CP-AgentNet Architecture}
CP-AgentNet comprises a CP-agent and three distinct types of memory. The CP-agent includes multiple agents such as a strategy agent, an observer agent, and a node agent, along with a programming assistant, each assigned specific roles to perform tasks. These agents are empowered by the LLM, each using tailored prompts suited to their individual roles. 
\\
\textbf{Strategy Agent.} The strategy agent is responsible for developing and refining strategies. It advises the node agent on selecting appropriate actions based on trajectory data and environmental changes and suggests methods to escape from suboptimal actions. \\
\textbf{Node Agent.} The primary role of the node agent is to determine the actions. Although the node agent generally adheres to the strategy, it relies on the observer agent for assistance in making the final decisions. The node agent adapts its action every $T$ slots, a period which we refer to as the \textit{query period}. Once established, the action is maintained for the duration of the query period.\\
\textbf{Observer Agent.} The observer agent diligently monitors the environment, focusing on the convergence of the action, any environmental changes, and notable occurrences. 
%Convergence is determined when Eq.~\ref{eq:conv} is satisfied\cite{convergence}. 
%\begin{equation}
%\frac{\sum_{i-9}^{i-5}(x_i) - \sum_{i-4}^{i}(x_i)}{\sum_{i-4}^{i}(x_i)} < \delta
%\label{eq:conv}
%\end{equation}
%where $x_i$ is the action at the $i$th round, and $\delta$ is a predefined small value. 
Environmental changes occur when new nodes join the network or existing nodes leave the network, or when other nodes change their parameters. Although the observer agent may not identify the specific changes, it is adept at detecting when changes occur. Notable occurrences can be any anomalies or outliers that the observer agent detects. For instance, if a specific slot is consistently overused or not used at all, the observer agent identifies this irregularity. Upon identifying such environmental changes or irregularities, the observer agent assists the node agent in adjusting its action accordingly.\\
\textbf{Programming Assistant.} Given that LLMs are not adept at complex mathematical tasks, we employ a programming assistant built upon LLMs. This agent supports other agents by solving mathematical problems using advanced programming skills. It transforms tasks into programming scripts and returns the executed results to the initiating agent. Additionally, this agent is integral to ASI, a process where the CP-Agent autonomously implements strategies.\\ 
\textbf{Memory.} CP-AgentNet employs three types of memories: strategy memory, episodic memory, and trajectory memory. As their names suggest, each memory type is dedicated to storing strategies, episodic data, and trajectories, respectively.

\subsection{CP-AgentNet Workflow}
The operation of CP-AgentNet consists of two stages: an offline stage and an online stage, as illustrated in Fig.~\ref{fig:CP-AgentNet}. In the offline stage, LLM-agents develop strategies that can be effectively utilized during the online stage. In the online stage, LLM-agents determine the optimal action based on these strategies.

\subsubsection{Generating Few-Shot Demonstrations}
To enable CP-AgentNet to learn communication strategies efficiently, we generate few-shot demonstrations leveraging simulation-based action-reward sampling and heterogeneous protocol representations. This process ensures that the LLM-agent can generalize across different network conditions and protocols with minimal training data. While our framework designed MAC protocol and TCP as use cases, applying it to other protocols would require generating separate few-shot demonstrations tailored to their characteristics.

\paragraph{Simulation-Based Demonstration Generation}
Instead of relying on large-scale datasets, we construct few-shot demonstrations through targeted simulation by leveraging in-context learning and generalization capabilities of LLM-agent. Given an environment characterized by state space $S$ and action space $A$, we sample a set of actions and derive the corresponding rewards from the simulation. Specifically, for each scenario, we:
\begin{itemize}
\item Sample $K$ actions, ${a_1, a_2, \dots, a_{K}}$, from the defined action space $A$.
\item Evaluate each action in the simulated environment to obtain the corresponding reward values, ${r_1, r_2, \dots, r_{K}}$.
\item Construct state-action-reward tuples $(s, a, r)$, where $s \in S$, $a \in A$, and $r$ is obtained from simulation feedback.
\item Store these tuples as part of the few-shot demonstration set $\mathcal{D}$ to be utilized in in-context learning for the LLM-agent.
\end{itemize}

\paragraph{Heterogeneous Demonstration Generation}
To ensure robustness across diverse network protocols, we generate demonstrations for both MAC and TCP protocols. Our framework constructs protocol-specific few-shot demonstrations that capture variations in communication behaviors.

For MAC protocols, we create total 4 demonstrations:
\begin{itemize}
\item Three protocol-specific demonstrations for \textbf{CSMA}, \textbf{TDMA}, and \textbf{ALOHA}.
\item One dynamic demonstration that adapts to changing network conditions, ensuring adaptability to real-world variations.
\end{itemize}

For TCP protocols, we generate total 3 demonstrations:
\begin{itemize}
\item Two protocol-specific demonstrations for \textbf{TCP Reno} and \textbf{TCP Vegas}.
\item One dynamic demonstration to account for variations in congestion and network dynamics.
\end{itemize}

%Each demonstration provides a structured reference for the LLM-agent to infer optimal strategies through in-context learning. The dynamic demonstrations further ensure that the model can generalize beyond predefined conditions, making CP-AgentNet robust in dynamic network environments. 

\begin{figure}
  \centering
  \centerline{\includegraphics[width=1\linewidth]{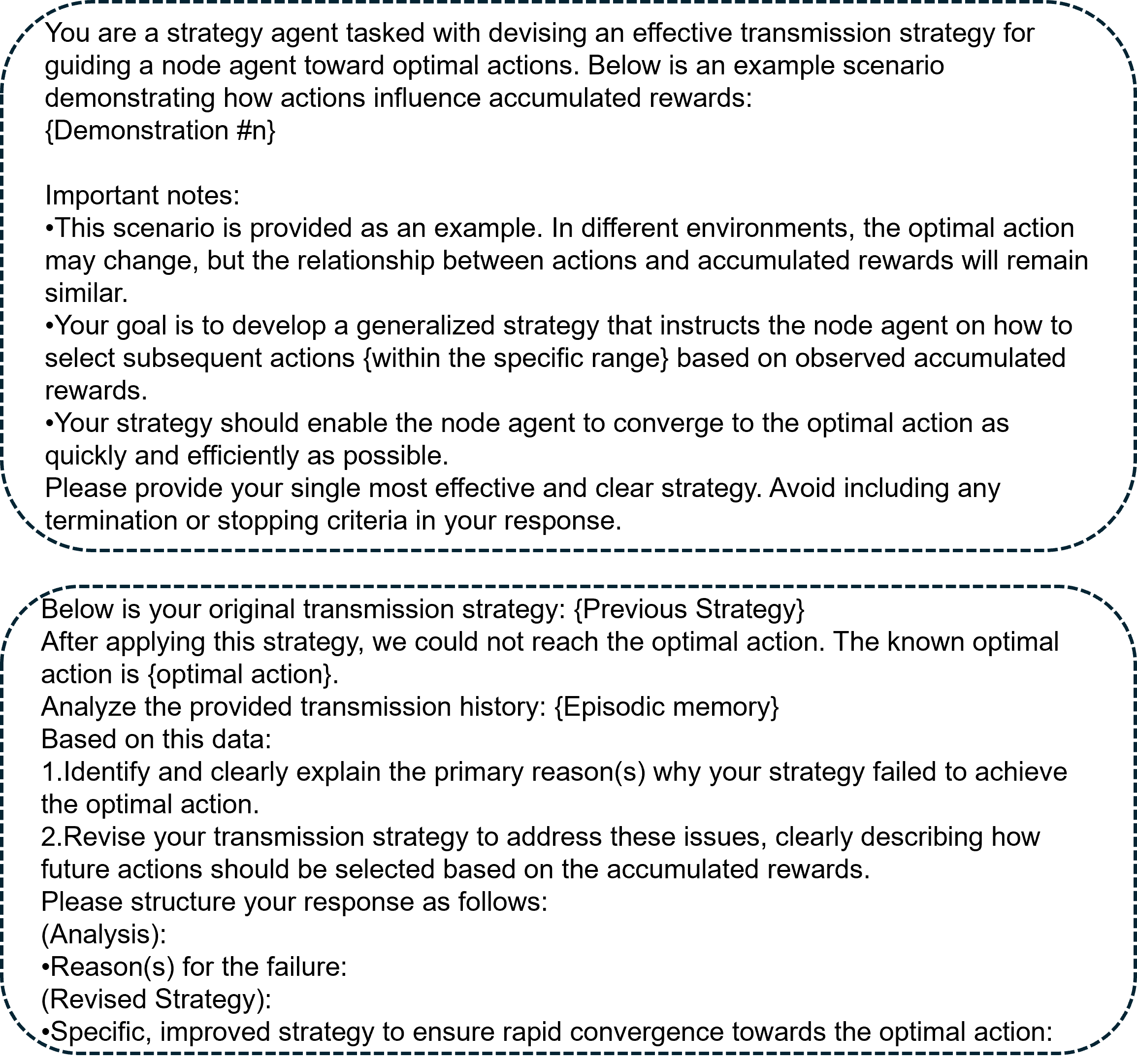}}
  \caption{Example prompts for strategy generation: initial strategy (top) and strategy refinement (bottom).}
  \label{fig:self-implementation}
\end{figure}

\subsubsection{Strategy generation} Strategy generation focuses solely on developing effective strategies based on the provided few-shot demonstrations. Rather than learning from a vast dataset, our approach enables LLM agents to infer strategies by analyzing various heterogeneous protocols and environmental scenarios. Through this process, the agent formulates strategies that can be effectively applied during the online adaptation stage. By distilling knowledge from limited demonstrations, the strategy generation phase ensures efficient adaptation while significantly reducing training time. 
We define a strategy $\pi$ that maps state space $S$ to action space $A$.
\begin{equation}
\pi: S \to A
\end{equation}
Given a set of few-shot demonstrations $\mathcal{D} = {(s_i, a_i, r_i, s'_i)}_{i=1}^{N}$, the optimal strategy is generated by maximizing the expected rewards over the sampled actions:

\begin{equation}
\pi^* = \arg\max_{\pi} \mathbb{E}_{(s, a, r) \sim \mathcal{D}} [R(s, a)]
\end{equation}
where $R(s, a)$ represents the reward function. Unlike traditional reinforcement learning, which requires extensive training, CP-AgentNet enables LLM-agents to leverage few-shot demonstrations and their generalization capabilities to construct an initial strategy.

\subsubsection{Refining Strategies} After generating an initial strategy, its effectiveness is evaluated by measuring the expected reward:

\begin{equation}
J(\pi) = \mathbb{E}_{s \sim P, a \sim \pi(s)} [R(s, a)]
\end{equation}
where $P$ represents the state distribution. If the obtained performance does not meet the predefined optimal threshold $J_{\text{opt}}$, i.e.,

\begin{equation}
J(\pi) < J_{\text{opt}}
\end{equation}
then the strategy undergoes a refinement process. This is accomplished through self-reflection\cite{reflexion}, enabling the LLM-agent to analyze its past decisions and adjust accordingly. The refinement process updates the strategy iteratively:

\begin{equation}
\pi^{(t+1)} = \pi^{(t)} + \lambda \nabla J(\pi^{(t)})
\end{equation}
where $t$ represents the iteration index and $\lambda$ is the step size for adjusting the policy. The process continues until the strategy reaches the desired performance:

\begin{equation}
J(\pi^{(t)}) \geq J_{\text{opt}} \quad \text{or} \quad t = N_{\max}
\end{equation}
By iteratively refining the strategy, CP-AgentNet ensures that suboptimal policies are corrected, allowing for efficient adaptation to dynamic environments. The refinement step enables the agent to incorporate feedback, adjust its decision-making process, and improve strategy performance without requiring extensive online learning. An example prompt for strategy generation is illustrated in Fig.~\ref{fig:self-implementation}.

\subsubsection{Online Adaptation} Once strategies are learned in the offline stage, the LLM-agent dynamically selects actions during the online stage while adapting to environmental variations. This adaptation is supported by the observer agent, which monitors the node agent's actions, evaluates environmental conditions, and assists in decision-making.

The node agent determines its action $a_t$ using the learned strategy $\pi^*$, with minor perturbations to allow adaptability:

\begin{equation}
a_t = \pi^*(s_t) + \omega_t, \quad \omega_t \sim \mathcal{N}(0, \sigma^2)
\end{equation}
Here, $\omega_t$ represents a small stochastic perturbation sampled from a normal distribution with zero mean and variance $\sigma^2$, ensuring that the agent remains adaptable to slight environmental fluctuations.
The observer agent assesses convergence, environmental changes, and notable occurrences.

\begin{equation}
a_t' = \pi^*(s_t, O_t)
\label{eq:node_action}
\end{equation}
where $O_t$ represents real-time feedback provided by the observer agent. Instead of directly modifying the action with an additive factor, the observer agent adjusts the decision-making process by providing contextual insights to the LLM-agent, ensuring robust adaptation to environmental dynamics. Notably, in the online stage, the strategy agent and episodic memory are inactive, as the system operates solely based on the pre-learned strategies and real-time observer feedback.

\begin{figure}
  \centering
  \centerline{\includegraphics[width=1\linewidth]{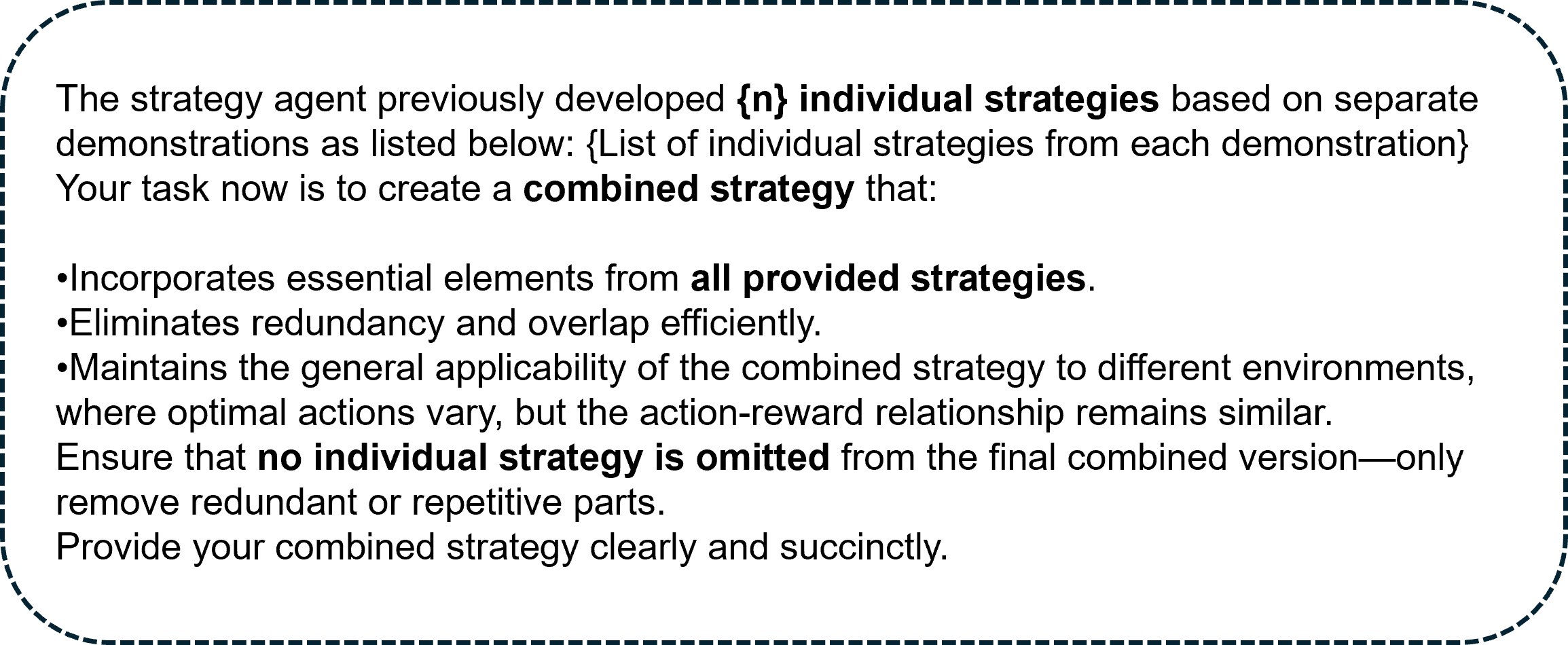}}
  \caption{An example prompt of PSA.}
  \label{fig:PSA}
\end{figure}

\begin{figure}
  \centering
  \centerline{\includegraphics[width=1\linewidth]{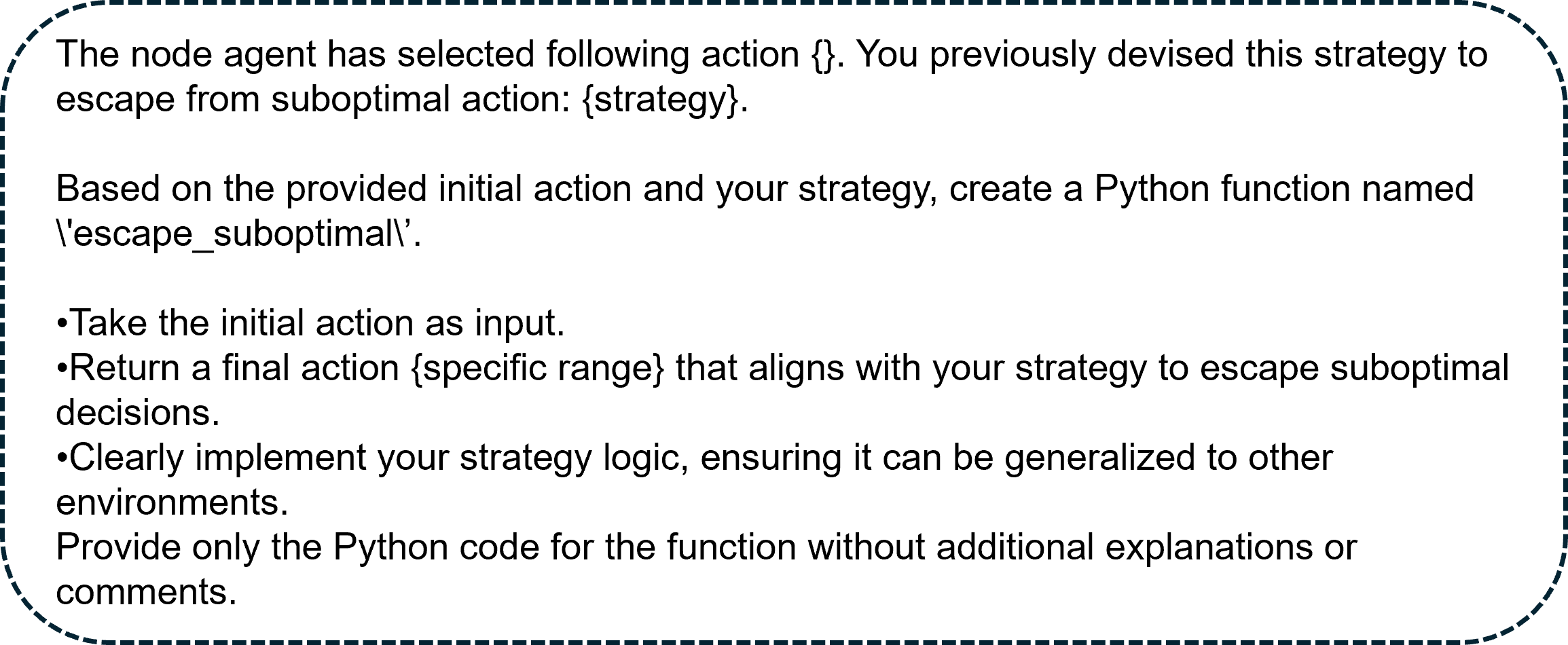}}
  \caption{An example prompt of ASI.}
  \label{fig:ASI}
\end{figure}

\begin{figure}[t]
  \centering
  \centerline{\includegraphics[width=1\linewidth]{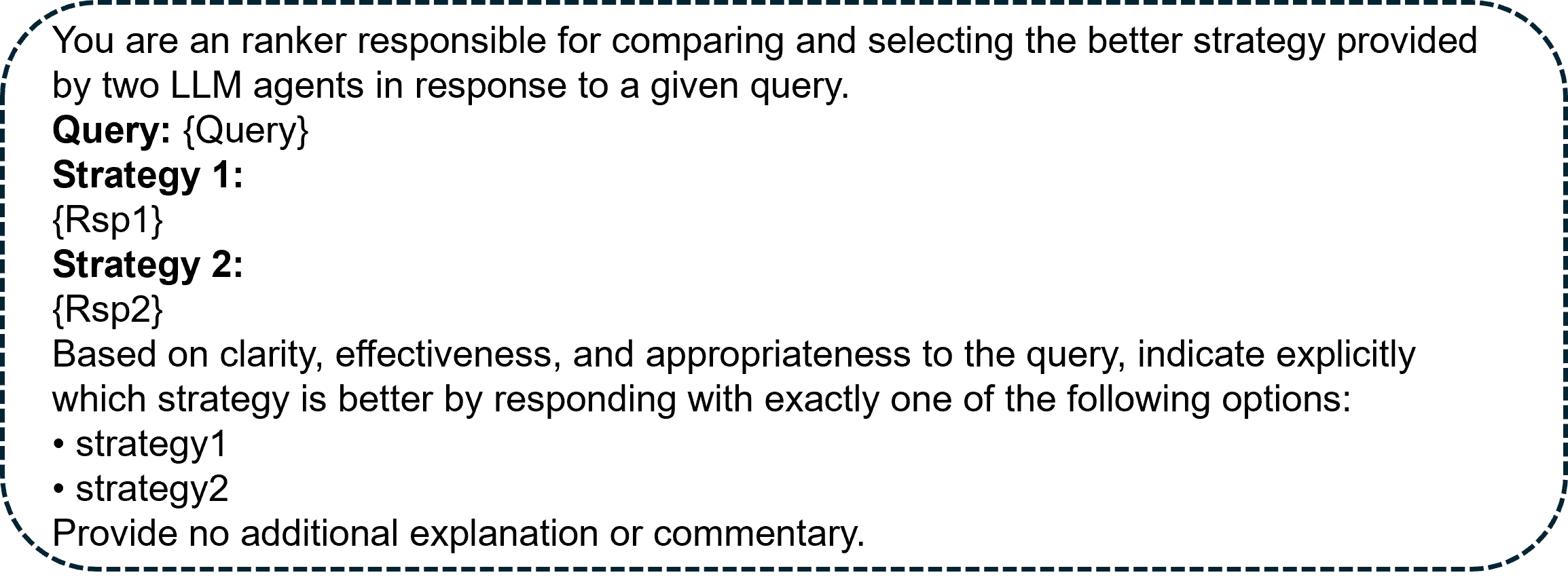}}
  \caption{An example prompt of LLM ranker.}
  \label{fig:LLM Ranker}
\end{figure}

\subsection{Autonomous Design}\label{Autonomous Design}

To achieve fully autonomous protocol design, CP-AgentNet must independently detect the causes of suboptimal performance and generate effective solutions without relying on explicit optimization methods, such as gradient descent. Instead of using user-provided feedback, CP-AgentNet autonomously refines its strategies based on its internal assessment of performance. If suboptimal performance is identified, the system dynamically expands its strategy set by incorporating a new strategy while ensuring efficiency and avoiding redundancy. The strategy agent proactively removes obsolete or conflicting strategies, maintaining a concise and optimal strategy set:

\begin{equation}
\Pi^{(t+1)} = \left(\Pi^{(t)} \cup \{\pi_{\text{new}}\} \right) \setminus \{\pi_{\text{obsolete}}\}
\end{equation}
This iterative approach, termed Progressive Strategy Adaptation (PSA), involves dynamically adapting strategies by continuously refining, integrating, and augmenting them. Unlike traditional approaches that depend on continuous weight updates, PSA facilitates progressive and context-aware strategy adaptation, resulting in improved autonomy and adaptability.

In addition, since the LLM's responses can be unpredictable, CP-AgentNet requires an autonomous mechanism to implement the generated strategies without human intervention. To address this, we introduce ASI, where the LLM generates executable functions that are reused throughout operation. This ensures efficient implementation without frequent re-querying or human involvement. For example, if the strategy agent suggests the $\epsilon$-greedy algorithm for an exploration-exploitation trade-off, the node agent implements this with the assistance of a programming assistant, which translates it into a function that operates autonomously at runtime. Example prompts for PSA and ASI are depicted in Fig.~\ref{fig:PSA} and Fig.~\ref{fig:ASI}, respectively.

\subsection{Stability of CP-AgentNet}
Despite their advantages, LLMs' inherently inconsistent outputs can compromise the stability of systems like CP-AgentNet. This inconsistency arises from their fundamental mechanism of predicting the next token’s probability distribution, even at zero temperature. Such unpredictability can degrade CP-AgentNet's performance or lead to operational failures. We have identified two primary sources of inconsistency: (i) compilation errors from the programming assistant and (ii) decision-making errors from LLM agents. To mitigate these issues, we introduce a compilation checker and an LLM ranker.

The programming assistant generates functions rather than complete scripts, which can lead to syntax errors or mismatched input parameters. To address function-level errors, we employ syntax checking, while input verification ensures parameter consistency. If a compilation error occurs, the assistant iteratively re-queries the LLM until a correct function is generated. While this process may be time-consuming, it remains feasible as it occurs during the non-time-sensitive offline stage.

Another critical challenge in CP-AgentNet is the inconsistency in decision-making by the CP-agent. To address this, we implement an LLM ranker \cite{ranker}, inspired by \textit{The Wisdom of Crowds} \cite{wisdom}. %As illustrated in Fig.~\ref{fig:ranker},
The LLM-agent queries the LLM twice, reversing the order of inputs to mitigate the influence of input sequencing on response generation. Both responses are then evaluated by another LLM acting as a judge, which selects the most optimal strategy autonomously.

\paragraph{Querying the LLM with Different Input Orders}
Let $Q$ be the original query from the CP-agent. To reduce bias from input ordering, we construct two variations:
\begin{equation}
Q_1 = Q, \quad Q_2 = \text{ReverseOrder}(Q).
\end{equation}
For each query, the LLM generates a response:
\begin{equation}
R_1 = \text{LLM}(Q_1), \quad R_2 = \text{LLM}(Q_2),
\end{equation}
where $R_i$ represents the strategy proposed by the LLM.

\paragraph{Selecting the Best Strategy Using Another LLM as a Judge}
Instead of manually defining evaluation criteria, we employ another LLM as a strategy judge $J_{\text{LLM}}$, which compares the two strategies and selects the best one:
\begin{equation}
R^* = J_{\text{LLM}}(R_1, R_2).
\end{equation}
where $J_{\text{LLM}}(R_1, R_2)$ outputs the preferred response based on its internal reasoning. The selection is entirely autonomous, relying on the judge LLM's ability to infer which response is more optimal. An example of LLM ranker is depicted in Fig.~\ref{fig:LLM Ranker}.

\paragraph{Feasibility in Offline Processing}
While the LLM ranker significantly improves decision consistency and overall reliability, it also increases processing time. However, since the LLM ranker is primarily utilized during the offline stage—where time constraints are less stringent—its impact on real-time performance is minimal. Specifically, the ranker is applied during the formulation of transmission strategies rather than during the time-sensitive online stage. The effectiveness of the LLM ranker will be demonstrated in Section \ref{Evaluation}.
\section{Use case protocols with CP-AgentNet}
In this section, we describe how we tailor the CP-AgentNet framework to design specific protocols. As use cases, we develop LLMA, a multiple access protocol, and CPTCP, a transmission control protocol, both designed for heterogeneous environments using CP-AgentNet.

\subsection{LLMA}
\subsubsection{Heterogeneous networks}
We consider heterogeneous networks where multiple nodes transmit data packets through a shared channel, fundamentally adhering to the assumptions outlined in DLMA \cite{DLMA}. We assume that the time slots are synchronized among these nodes, each using different MAC protocols. Each node initiates its data packet transmission at the beginning of a slot and completes it within the same slot. Every node consistently has a data packet to transmit. If more than one node transmits a data packet using the same channel simultaneously, a collision occurs, rendering those transmissions unsuccessful. 
We consider heterogeneous network scenarios where at least one ``heteronode'' coexists with slotted ALOHA or TDMA nodes. A slotted ALOHA node transmits a data packet with a fixed probability $q$ in each time slot, while a TDMA node transmits data packets at $X$ specific slots out of a frame composed of ten slots. The term ``heteronode'' refers to nodes such as AWARE node, DLMA node, or LLMA node. The AWARE node, considered an ideal solution, has complete knowledge of the environment, including the number of slotted ALOHA and TDMA nodes and their parameters, represented by $q$ and $X$. Consequently, our goal is to enable the LLMA node to perform as closely as possible to the AWARE node. A DLMA node utilizes the DLMA protocol, which is developed using a DRL method, to make transmission decisions. It can listen to the channel and observe whether other nodes' transmissions are successful or if the channel is idle. Similarly, an LLMA node employs the LLMA protocol, which is developed through our CP-AgentNet framework. All assumptions for LLMA are the same as for DLMA unless specified otherwise. 

Additionally, we consider more complex scenarios where the heteronode coexists not only with TDMA and ALOHA nodes but also with CSMA (carrier-sense multiple access), EB (exponential backoff window)-ALOHA, and FB (fixed window)-ALOHA nodes. The CSMA node generates a random value of $w$ within the range [0, $W$-1], and senses the channel is busy or not. If it is sensed as idle, $w$ is decreased by one, otherwise it is frozen. If this value reaches to zero, the CSMA node transmits data. Only the CSMA node has the capability to sense the channel, and the sensing time is negligible compared to the packet length. An FW-ALOHA node generates a random value of $w$ within the range [0, $W$-1] after transmission, it then waits for $w$ slots for the next transmission. An EB-ALOHA node's operation is basically the same as that of an FW-ALOHA node. However, an EB-ALOHA node doubles its window size upon a collision with other nodes, up to a maximum window size of $2^{m}W$. In these scenarios, only DLMA and LLMA nodes are used as heteronodes, as we do not derive the ideal behavior for the AWARE node.

\subsubsection{Objective function of heteronode}
We set the objective for the heteronode to maximize $\alpha$-fairness throughput, where $alpha=1$, rather than focusing solely on total throughput. 
$\alpha$-fair function is defined as Eq.~\ref{eq:alpha_fairness}.
\begin{equation}
    g_{\alpha}(x)= 
\begin{cases}
    \log{x},& \text{if } \alpha = 1\\
    \frac{x^{1-\alpha}}{1-\alpha},   & \text{otherwise}
\end{cases}
\label{eq:alpha_fairness}
\end{equation}
\begin{equation}
f(x_1, ..., x_N) = \sum_{i=1}^{N}g_{\alpha}(x_i)
\label{eq:sum alpha}
\end{equation}
where $N$ is the number of nodes, $x_i$ represents the $\alpha$-fair throughput of note $i$.
Our goal is to maximize $f(\vec{x})$, the sum of $\alpha$-fair function across all nodes. Setting $\alpha$ to zero corresponds to maximizing total throughput. We set our objective to maximize $\alpha$-fairness specifically when $\alpha$ is 1.

\subsubsection{LLMA design}
The goal of LLMA is to optimize the action of a node agent based on strategies and real-time environmental feedback. In this context, the action determines the likelihood of transmitting in each time slot, setting the transmission probability for each slot. At time step $t$, the action $a_t$ is selected from continuous range $[0,1]$. The observation space at time $t$, denoted as $o_t$, includes `S' for `SUCCESSFUL', `C' for `COLLIDED', or `I' for `IDLE'. `S' indicates that only one node transmits a data packet at time $t$, `C' signifies that more than one node transmits data packets in the same time slot, and `I' means that none of the nodes transmit a data packet. It is important to note that this observation $o_t$ is distinct from the observer feedback $O_t$ in Eq.~\ref{eq:node_action}, where $O_t$ represents higher-level real-time feedback used by the LLM agent to adjust its action. In this setup, the observation space directly serves as the state space $S$. Rewards from the environment, given after an action, are defined as follows: $r_{t+1} =1$ if $o_t$ is `S', and $r_{t+1} =0$ if $o_t$ is either `C' or `I'. Assuming there are $N$ nodes in the network, the node agent receives an $N$ dimension vector of rewards from the environment. Each element of the vector represents the transmission result of one particular node. The actions taken by the node agent and the rewards from the environment are essentially the same as those in DLMA\cite{DLMA}.

The node agent adapts its action every $T$ slots, a period which we refer to as the query period. During this time, the accumulated actions and rewards, along with the observations, are stored in the trajectory memory. Before the node agent adapts the action at the end of the query period, the observer agent monitors three aspects: the convergence of the plan, any environmental changes, and any notable occurrences. 
With assistance from the programming assistant, if the observer agent determines that the transmission has converged, the node agent ceases adaptation and implements the strategy to escape from suboptimal as described in Section \ref{Autonomous Design}. If the observer agent detects any environmental changes, it notifies the node agent to adjust the plan accordingly. When coexisting with a TDMA node, if the LLMA node's observer agent notices that specific slots are consistently used, it advises the node agent to avoid these slots. Through collaboration with the observer agent and the programming assistant, the node agent decides the action for each slot.

\subsection{CPTCP}
\subsubsection{Heterogeneous TCPs} We consider heterogeneous TCP scenarios in which CPTCP can coexist with different TCP algorithms. There are mainly two types of TCP algorithms: loss-based and delay-based TCP. Fairness issues arise when a loss-based TCP flow coexists with a delay-based TCP flow, as the loss-based TCP reactively reduces its congestion window size, while the delay-based TCP proactively reduces it \cite{fairnessTCP}. Although previous studies have attempted to overcome this problem, our goal is to mitigate it by leveraging the in-context learning capability of the LLM, ultimately enabling CPTCP to effectively coexist with both loss-based and delay-based TCP flows without the need for hand-crafted design requiring human effort. We employ TCP Reno and TCP Vegas\cite{vegas} as representatives for loss-based and delay-based TCP, respectively.

\subsubsection{Objective function of CPTCP} 
Since we focus on fairness when coexisting with different types of TCP flows, our objective is to maximize Jain's fairness index, which is defined in Eq. \ref{eq:jain fairness}.
\begin{equation}
f(x_1, ..., x_N) = \frac{(\sum_{i=1}^{N}x_i)^2}{N\sum_{i=1}^{N}x_i^2}
\label{eq:jain fairness}
\end{equation}
where $x_i$ is the throughput of $i$th flow and N is the number of flows. In a completely fair scenario, $f(\vec{x})=1$. Conversely, if a single TCP flow monopolizes the link, $f(\vec{x})=\frac{1}{N}$. Thus, the closer $f(\vec{x})$ is to 1, the greater the fairness; conversely, the closer $f(\vec{x})$ is to $\frac{1}{N}$, the lower the fairness.

\subsubsection{CPTCP Design} The design procedure for CPTCP follows the same foundational principles as LLMA, with modifications tailored to its specific objectives. Accordingly, the objective of CPTCP is to optimize the action of a node agent based on strategies and real-time environmental feedback. In this context, the action determines the congestion window. At time step \( t \), the action \( a_t \) represents the congestion window size, which is a positive integer within a predefined range. Specifically, we define the action space as $
a_t \in [1, C_{\max}]$, where \( C_{\max} \) is the maximum congestion window size, determined by factors such as the receiver’s advertised window and network conditions. While the theoretical limit can be large in high-bandwidth networks, practical implementations impose constraints based on these factors. While the theoretical maximum congestion window can be large in high-bandwidth networks, practical implementations impose constraints based on these factors. The observation space at time \( t \), denoted as \( o_t \), consists of two key metrics: \( a \), the number of received acknowledgments (ACKs), and \( r \), the round-trip time (RTT), forming a vector representation \( o_t = [a, r] \).  The environment provides rewards following each action, $r_{t+1} = \log(a) - \beta r$. Here, the logarithmic function promotes fairness in throughput allocation, while \( \beta \) serves as a tunable parameter balancing throughput and RTT trade-offs. If the observer agent detects any environmental changes, it notifies the node agent, which then adapts the action accordingly. The final decision is made based on the updated conditions, ensuring an optimized response to dynamic network environments.

\section{Evaluation}\label{Evaluation}

\subsection{Simulation Setup}
To evaluate the effectiveness of CP-AgentNet, we conducted experiments with LLMA and CPTCP. CP-Agent is powered by GPT-4-o, with the temperature value set to zero to minimize inconsistencies. 
\subsubsection{LLMA}
We focus on heterogeneous networks incorporating ALOHA and TDMA protocols, alongside heteronodes such as AWARE nodes, DLMA nodes, and LLMA nodes. Since our MAC protocol evaluation already includes direct quantitative comparisons against both an ideal baseline and a representative DRL-based method, additional comparisons against other methods are unnecessary to sufficiently validate the performance of CP-AgentNet. Specifically, within these networks, only one type of heteronode coexists with nodes operating under TDMA or ALOHA protocols. Each frame consists of ten time slots, each lasting 1ms. Specifically for TDMA nodes, we consider a scenario where only one TDMA node utilizes slots 3 and 5, denoted as $X=2$. For ALOHA nodes, we set the transmission probability to $q=0.2$ except for the masssive nodes case. For a massive nodes scenario, $q$ is set to 0.02. Moreover, to demonstrate the robustness of LLMA nodes within complex heterogeneous networks, we conduct additional tests involving nodes that utilize CSMA, FW-ALOHA, and EB-ALOHA. We set $W$ to 4 for FW-ALOHA, and for CSMA and EB-ALOHA, we set window size $W$ to 2 with maximum backoff stage $m$ equal to 4 and 2, respectively. Following the setting in DLMA \cite{DLMA}, we use throughput as the performance metric in this paper. However, since the objective is not to maximize total throughput but to maximize $\alpha$-fairness throughput for heteronodes, we use RMSE as an additional metric to measure how closely the behavior approximates the ideal operation of the AWARE node.
\subsubsection{CPTCP} We focus on three TCP algorithms: TCP Reno, TCP Vegas\cite{vegas}, and CPTCP. 
We first evaluate the throughput of homogeneous TCP configurations, followed by assessments of various combinations involving different TCP algorithms. We utilize a network capacity of 1 MHz, with throughput and Jain's fairness index serving as the evaluation metrics. Since the primary goal of our CPTCP evaluation is to demonstrate its capability to coexist with diverse TCP variants, we benchmark CPTCP against representative loss-based (TCP Reno) and delay-based (TCP Vegas) algorithms.

\subsection{LLMA results}
We initially define various scenarios to assess the robustness of LLMA nodes, which are detailed in Table~\ref{tab:RMSE} along with their respective RMSE results. In the following, we will review these scenarios and analyze the corresponding outcomes.

%\begin{figure*}
%     \begin{subfigure}[b]{0.3\textwidth}
%         \centering
%         \includegraphics[width=\textwidth]{figs/results/H1_A1_T1/AWARE1_A1_T1.pdf}
%         \caption{AWARE node+1T+3A}
%         \label{fig:AWARE node coexisting with 3A+1T}
%     \end{subfigure}
%     \hfill
%     \begin{subfigure}[b]{0.3\textwidth}
%         \centering
%         \includegraphics[width=\textwidth]{figs/results/H1_A1_T1/DLMA1_A1_T1.pdf}
%         \caption{DLMA node+1T+3A}
%         \label{fig:Residual Block}
%     \end{subfigure}
%     \hfill
%     \begin{subfigure}[b]{0.3\textwidth}
%         \centering
%         \includegraphics[width=\textwidth]{figs/results/H1_A1_T1/LLMA1_A1_T1.pdf}
%         \caption{LLMA node+1T+3A}
%         \label{fig:Bottleneck Block}
%     \end{subfigure}
%    \caption{Throughput of single heteronode coexisting with three slotted ALOHA nodes and 1 TDMA node}
%    \label{fig:1H+3A+1T}
%\end{figure*}

\subsubsection{\textbf{Homogeneous scenario}}
\begin{figure}
     \begin{subfigure}[b]{0.15\textwidth}
         \centering
         %\includesvg[width=\columnwidth]{figs/results/Homo/LLMA2_edit.svg}
         \includegraphics[width=\columnwidth]{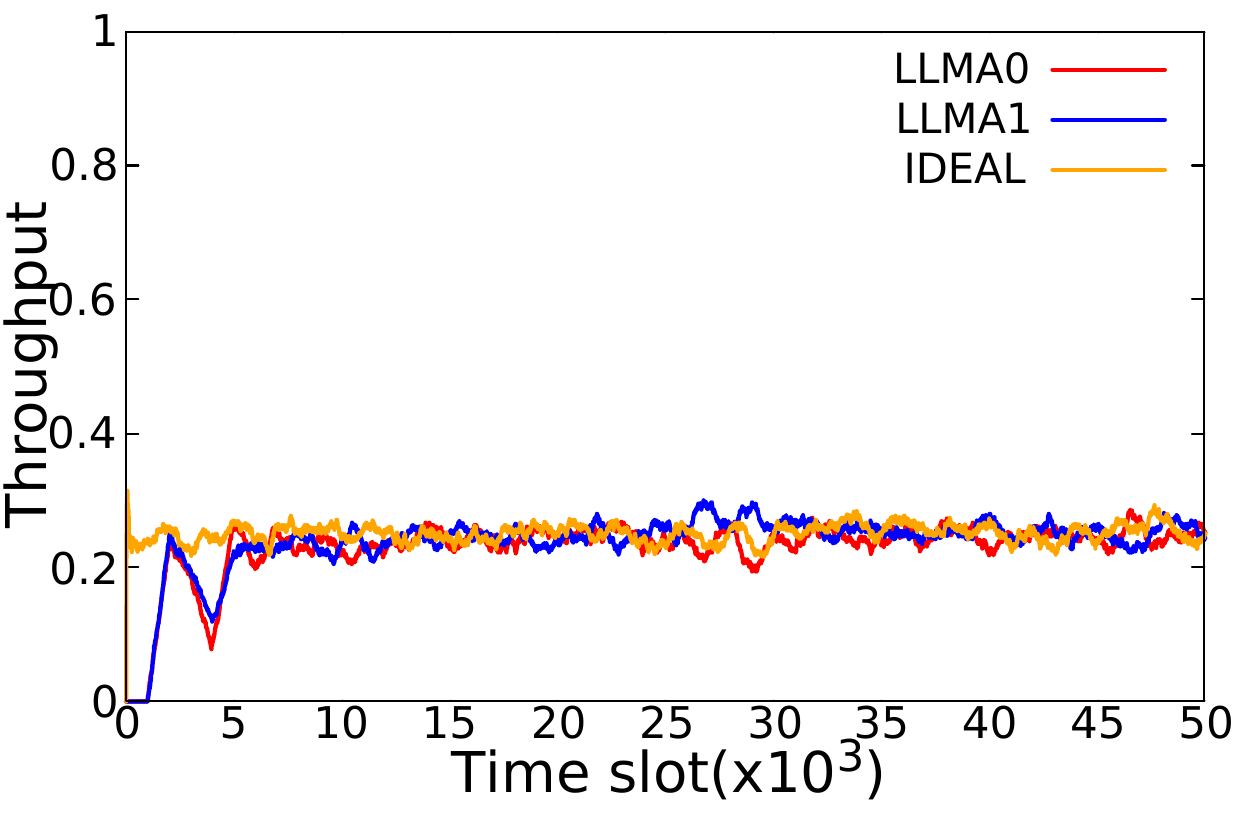}
         \caption{2 LLMA node}
         \label{fig:2 LLMA nodes}
     \end{subfigure}
     \hfill
     \begin{subfigure}[b]{0.15\textwidth}
         \centering
         %\includesvg[width=\columnwidth]{figs/results/Homo/LLMA4_edit.svg}
         \includegraphics[width=\columnwidth]{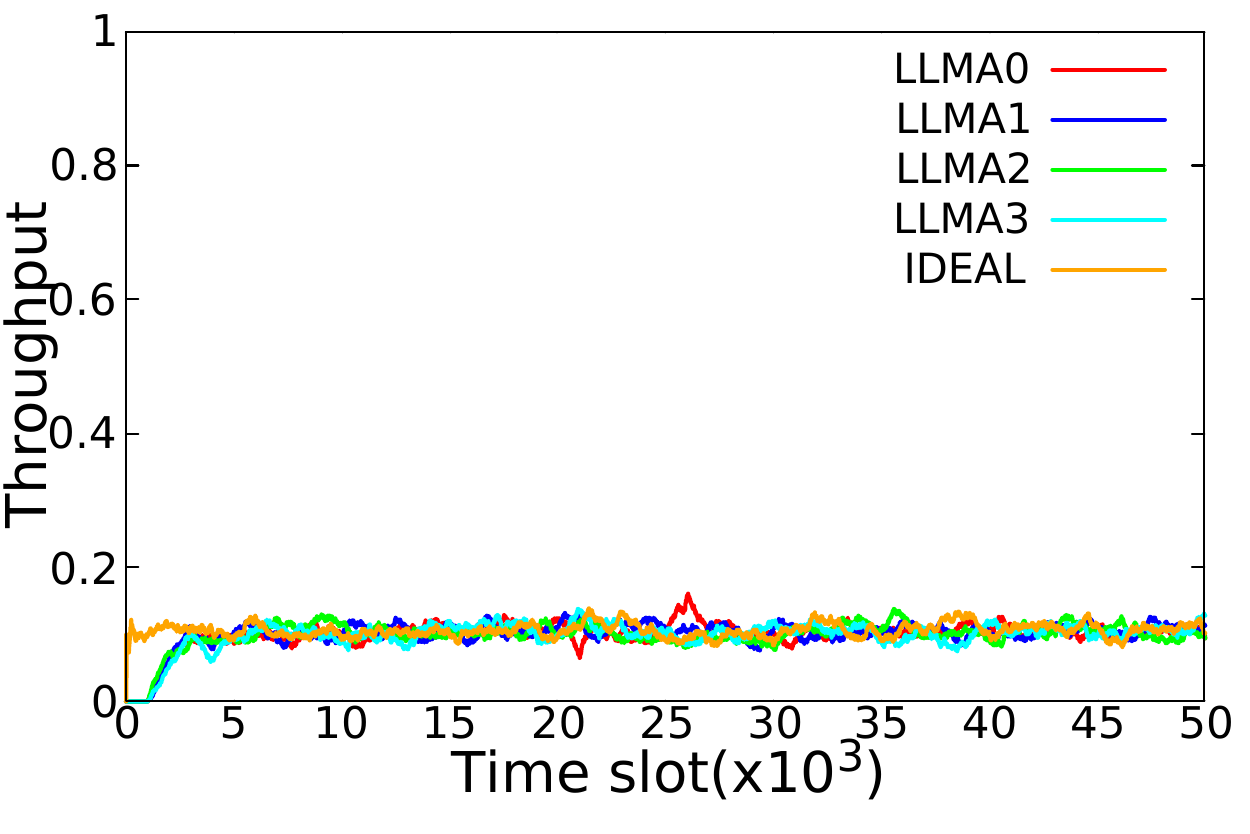}
         \caption{4 LLMA nodes}
         \label{fig:4 LLMA nodes}
     \end{subfigure}
     \hfill
     \begin{subfigure}[b]{0.15\textwidth}
         \centering
         %\includesvg[width=\columnwidth]{figs/results/Homo/LLMA8_edit.svg}
         \includegraphics[width=\columnwidth]{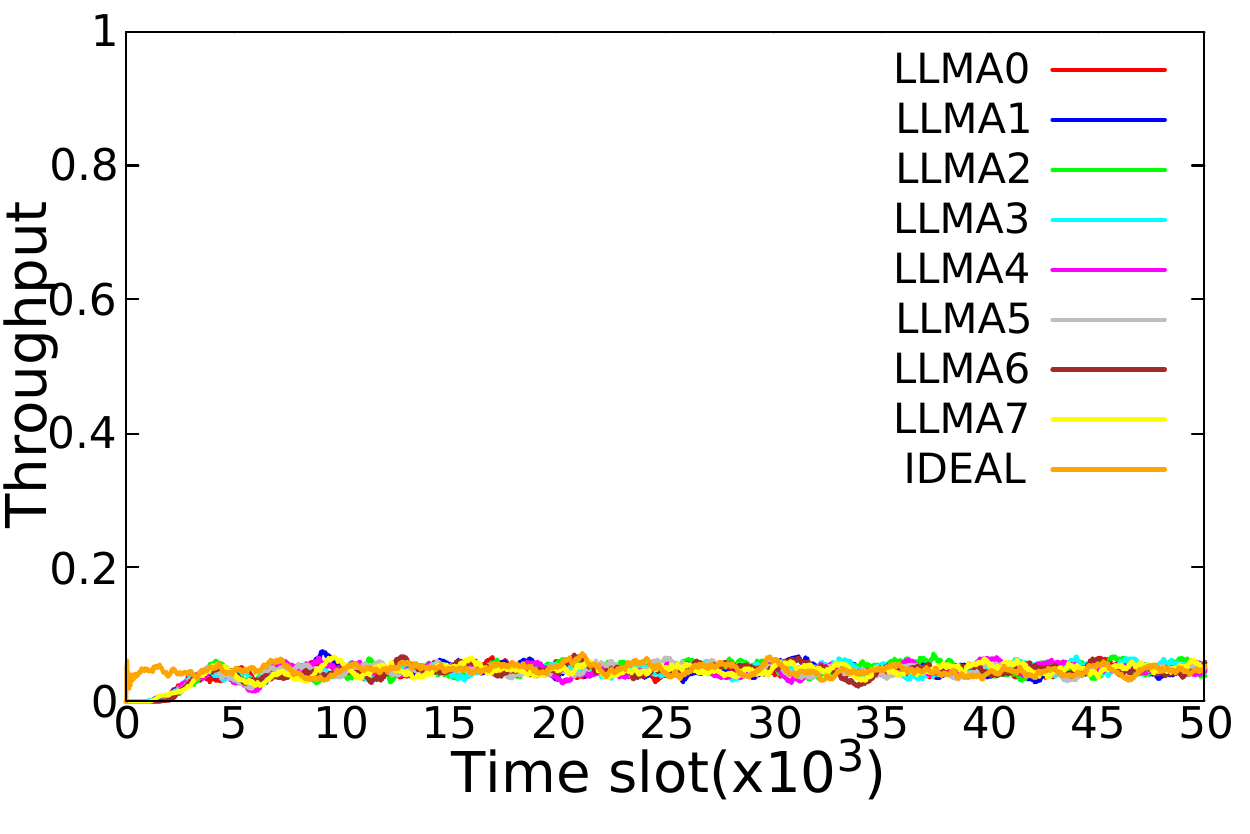}
         \caption{8 LLMA nodes}
         \label{fig:8 LLMA nodes}
     \end{subfigure}
     \setlength{\belowcaptionskip}{-10pt}
    \caption{LLMA nodes in homogeneous environment}
    \label{fig:homo}
\end{figure}

We evaluate multiple LLMA nodes in configurations of 2, 4, and 8, and compare the results with the ideal behavior of the AWARE node. After conducting separate simulations with AWARE nodes in the same configurations, we plot the throughput of one AWARE node alongside the LLMA nodes for comparison in Fig.~\ref{fig:homo}. The throughput of the LLMA nodes closely aligns with the ideal behavior, demonstrating that the emerging behavior of the LLMA nodes matches the optimal behavior.

\subsubsection{\textbf{Heterogeneous scenario}}
\begin{figure}
     \begin{subfigure}[b]{0.15\textwidth}
         \centering
         %\includesvg[width=\columnwidth]{figs/results/H1_A3_T1/AWARE1_A3_T1_edit.svg}
         \includegraphics[width=\columnwidth]{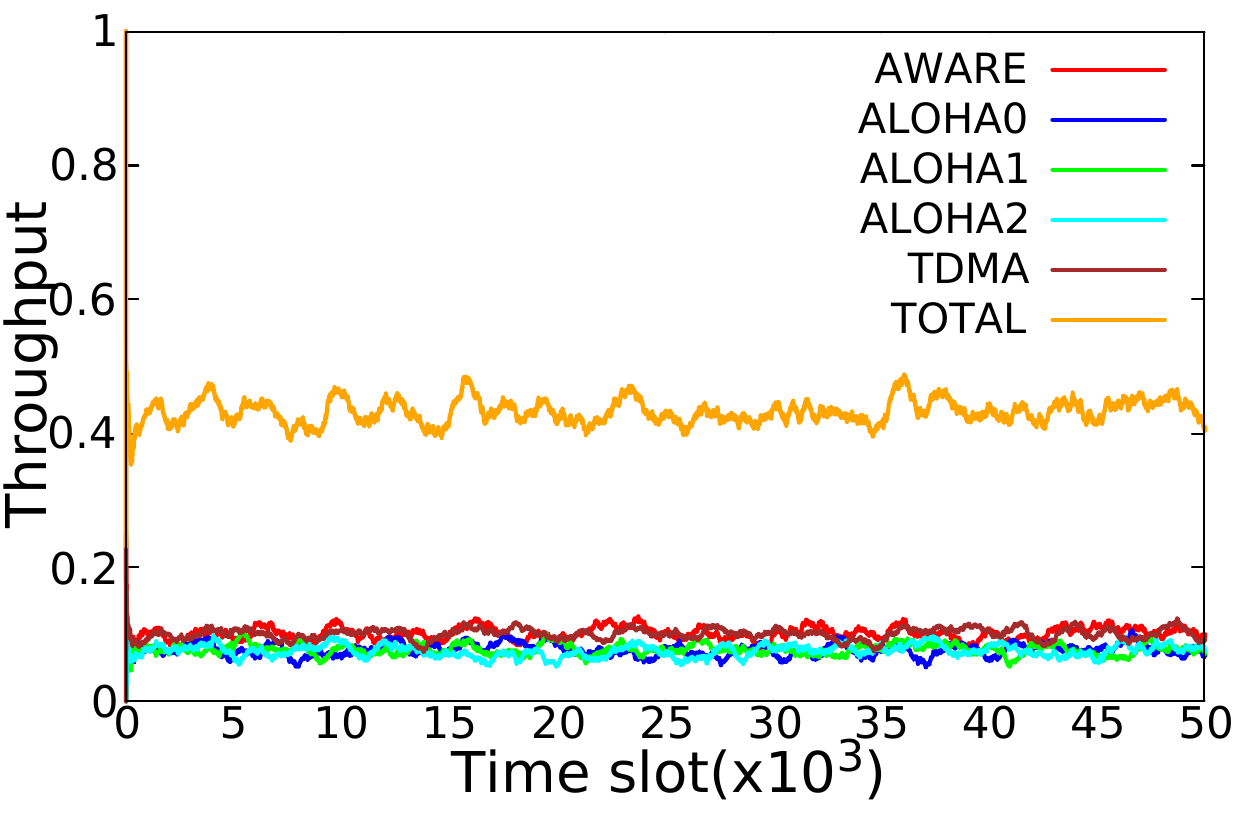}
         
         \caption{AWARE+1T+3A}
         \label{fig:AWARE node coexisting with 3A+1T}
     \end{subfigure}
     \hfill
     \begin{subfigure}[b]{0.15\textwidth}
         \centering
         %\includesvg[width=\columnwidth]{figs/results/H1_A3_T1/DLMA1_A3_T1_edit.svg}
         \includegraphics[width=\columnwidth]{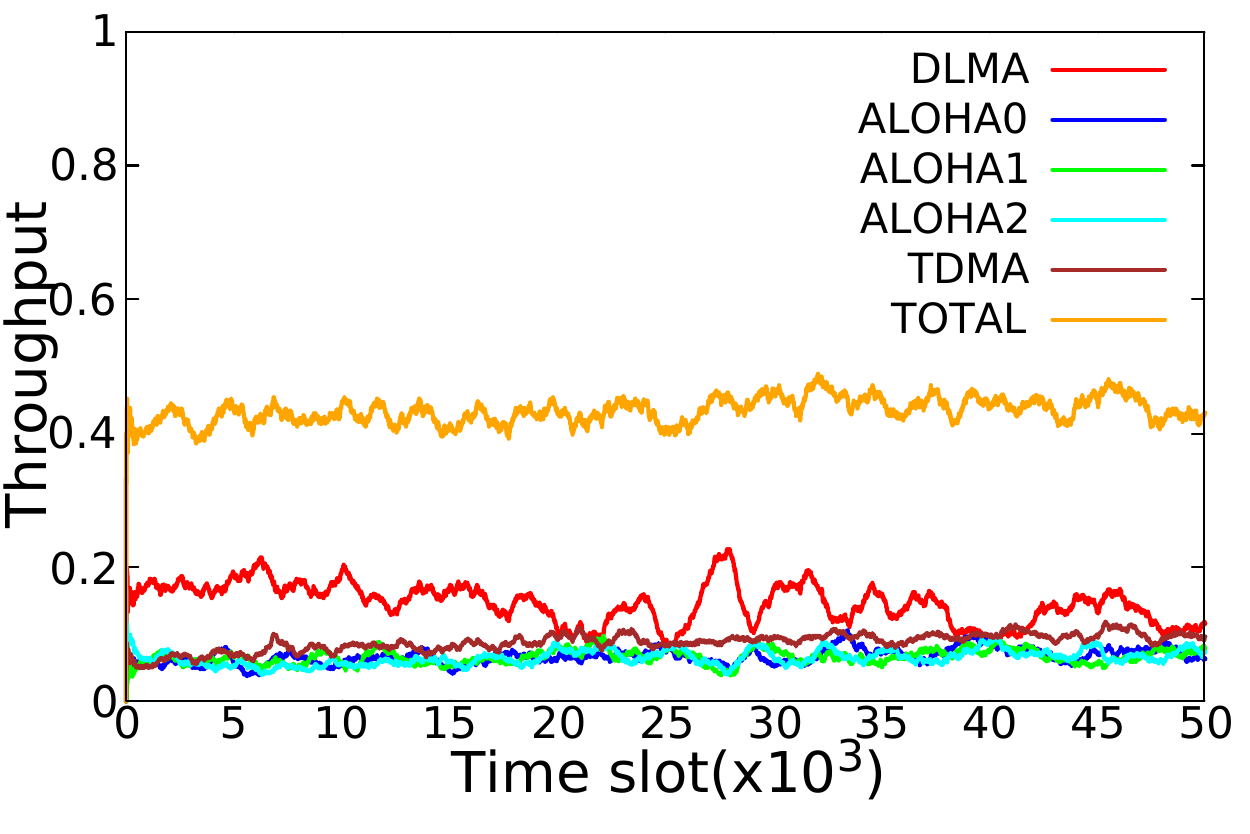}
         \caption{DLMA+1T+3A}
         \label{fig:DLMA node coexisting with 3A+1T}
     \end{subfigure}
     \hfill
     \begin{subfigure}[b]{0.15\textwidth}
         \centering
         %\includesvg[width=\columnwidth]{figs/results/H1_A3_T1/LLMA1_A3_T1_edit.svg}
         \includegraphics[width=\columnwidth]{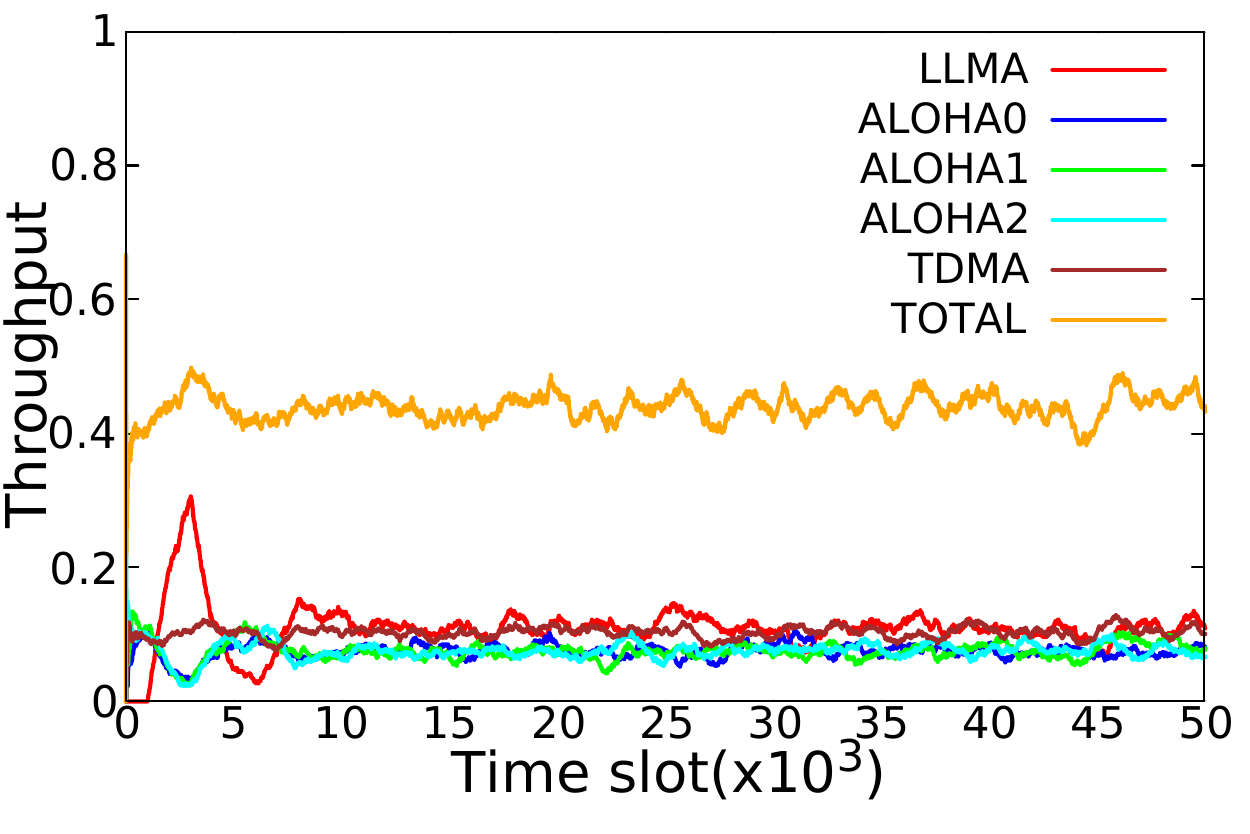}
         \caption{LLMA+1T+3A}
         \label{fig:LLMA node coexisting with 3A+1T}
     \end{subfigure}
     \setlength{\belowcaptionskip}{-10pt}
    \caption{Heteronode with 3 ALOHA and 1 TDMA nodes}
    \label{fig:1H+3A+1T}
\end{figure}

To demonstrate that LLMA nodes can coexist with nodes using different protocols, we explore various combinations involving TDMA, ALOHA nodes, and heteronodes. The illustrations in Fig.~\ref{fig:1H+3A+1T}, arranged from left to right, depict an AWARE node, a DLMA node, and an LLMA node. Although the LLMA node's behavior visually appears similar to that of the AWARE nodes, we used RMSE to objectively confirm this similarity. The results for all other scenarios are presented in the Table~\ref{tab:RMSE}. Except for one scenario (1H+1T), the RMSE values of the LLMA node are smaller than those of the DLMA node, indicating that the LLMA node's operation is closer to that of the AWARE node.
In scenarios where multiple heteronodes coexist with different nodes, the RMSE values of LLMA nodes are approximately 49.4\% and 6.8\% lower than those of DLMA nodes in the 2A+2H and 1T+2A+3H scenarios, respectively. In summary, our evaluations in heterogeneous scenarios, similar to those examined in \cite{DLMA}, demonstrate that the CP-AgentNet framework manages heterogeneous network environments more effectively than traditional DRL methods.

\begin{figure}[t!]
     \centering
     \begin{subfigure}[b]{0.15\textwidth}
         \centering
         %\includesvg[width=\columnwidth]{figs/results/Dynamic/AWARE_edit3.svg}
         \includegraphics[width=\columnwidth]{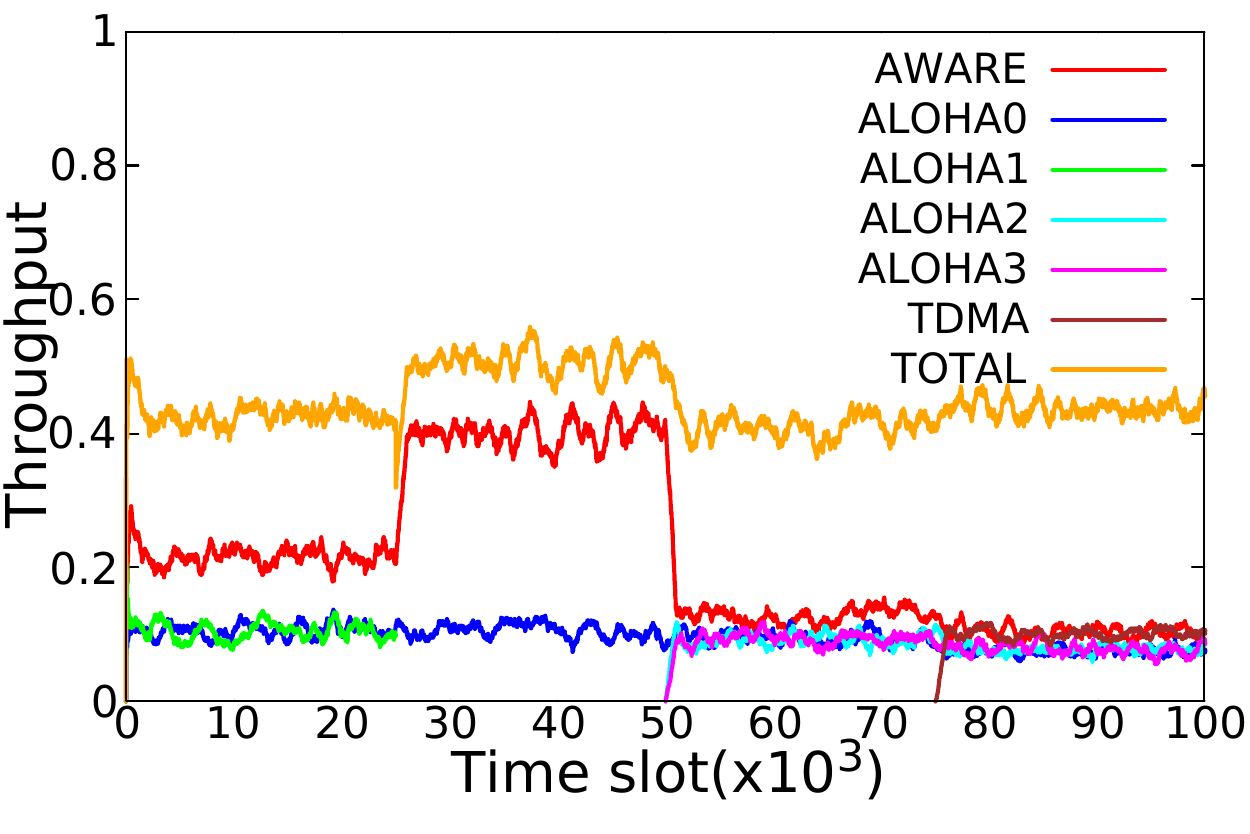}
         \caption{AWARE node}
         \label{fig:Basic Block}
     \end{subfigure}
     \hfill
     \begin{subfigure}[b]{0.15\textwidth}
         \centering
         %\includesvg[width=\columnwidth]{figs/results/Dynamic/DLMA_edit.svg}
         \includegraphics[width=\columnwidth]{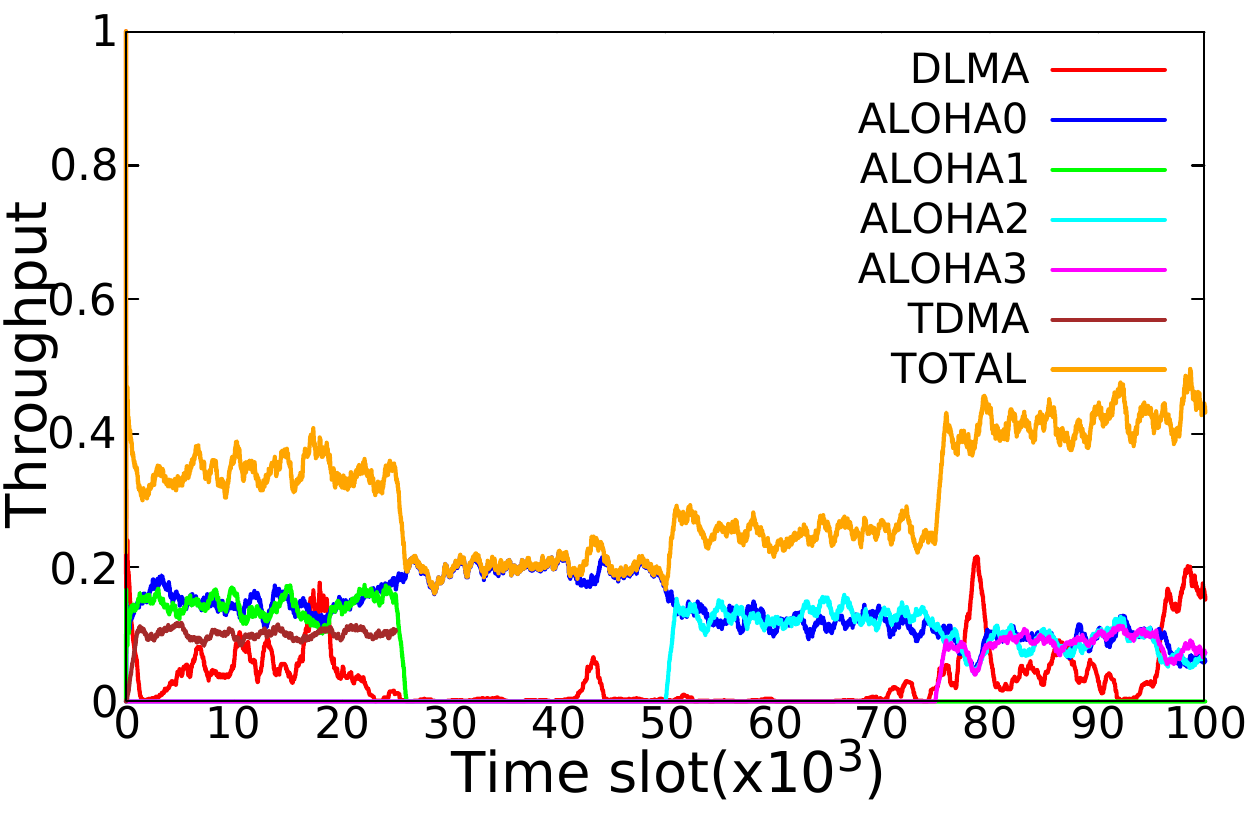}
         \caption{DLMA node}
         \label{fig:Residual Block}
     \end{subfigure}
     \hfill
     \begin{subfigure}[b]{0.15\textwidth}
         \centering
         %\includesvg[width=\columnwidth]{figs/results/Dynamic/LLMA_edit.svg}
         \includegraphics[width=\columnwidth]{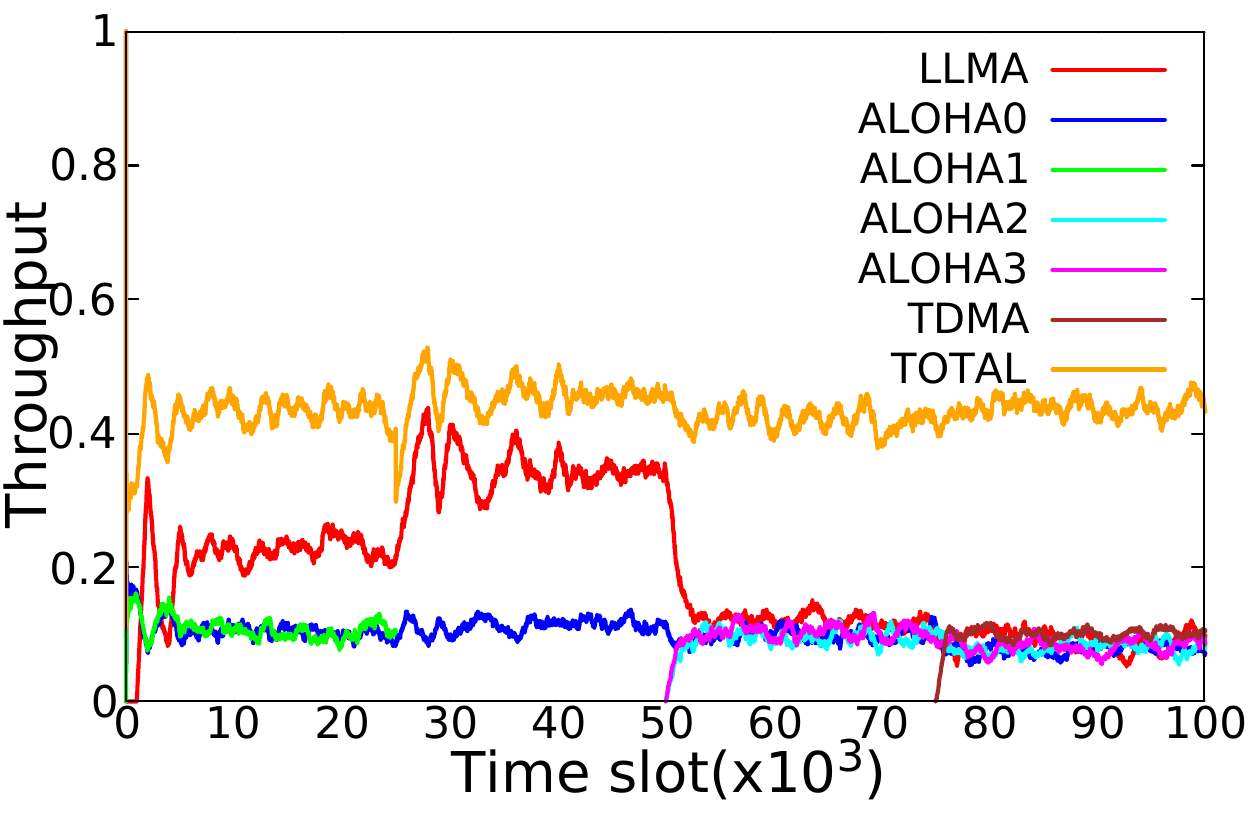}
         \caption{LLMA node}
         \label{fig:Bottleneck Block}
     \end{subfigure}     
    \setlength{\belowcaptionskip}{-10pt}
        \caption{Heteronode in a dynamic environment}
        \label{fig:single_dynamic}
\end{figure}

\subsubsection{\textbf{Complex Heterogeneous scenario}}
\begin{figure}
     \begin{subfigure}[b]{0.48\columnwidth}
         \centering
         \includegraphics[width=\columnwidth]{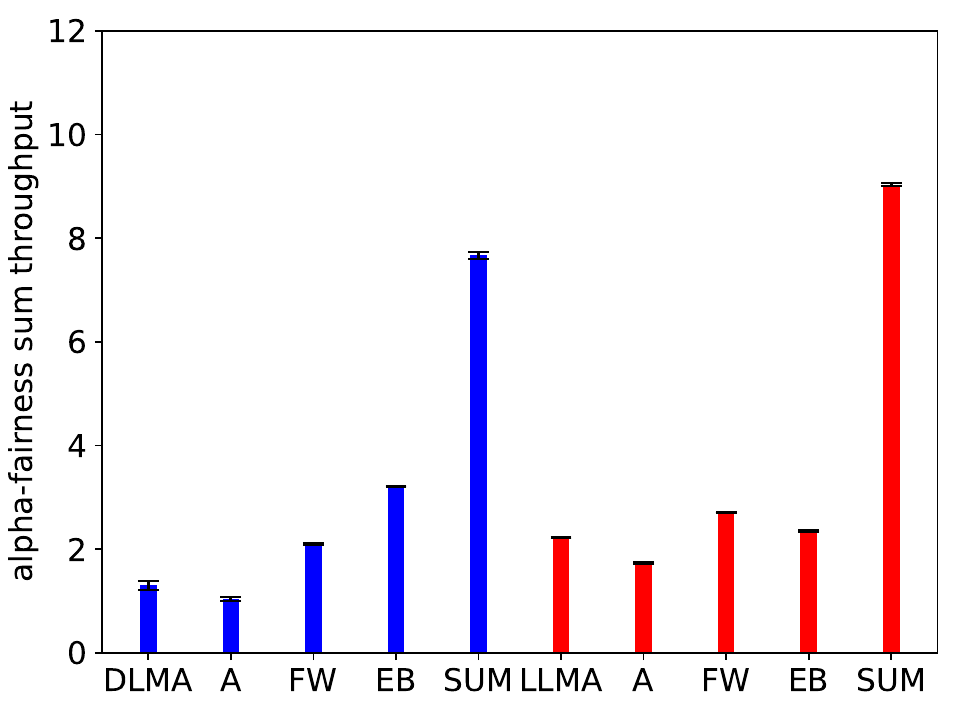}
         \caption{1H+1A+1EB+1FW}
         \label{fig:1H+1A+1EB+1FW}
     \end{subfigure}
     \hfill
     \begin{subfigure}[b]{0.48\columnwidth}
         \centering
         \includegraphics[width=\textwidth]{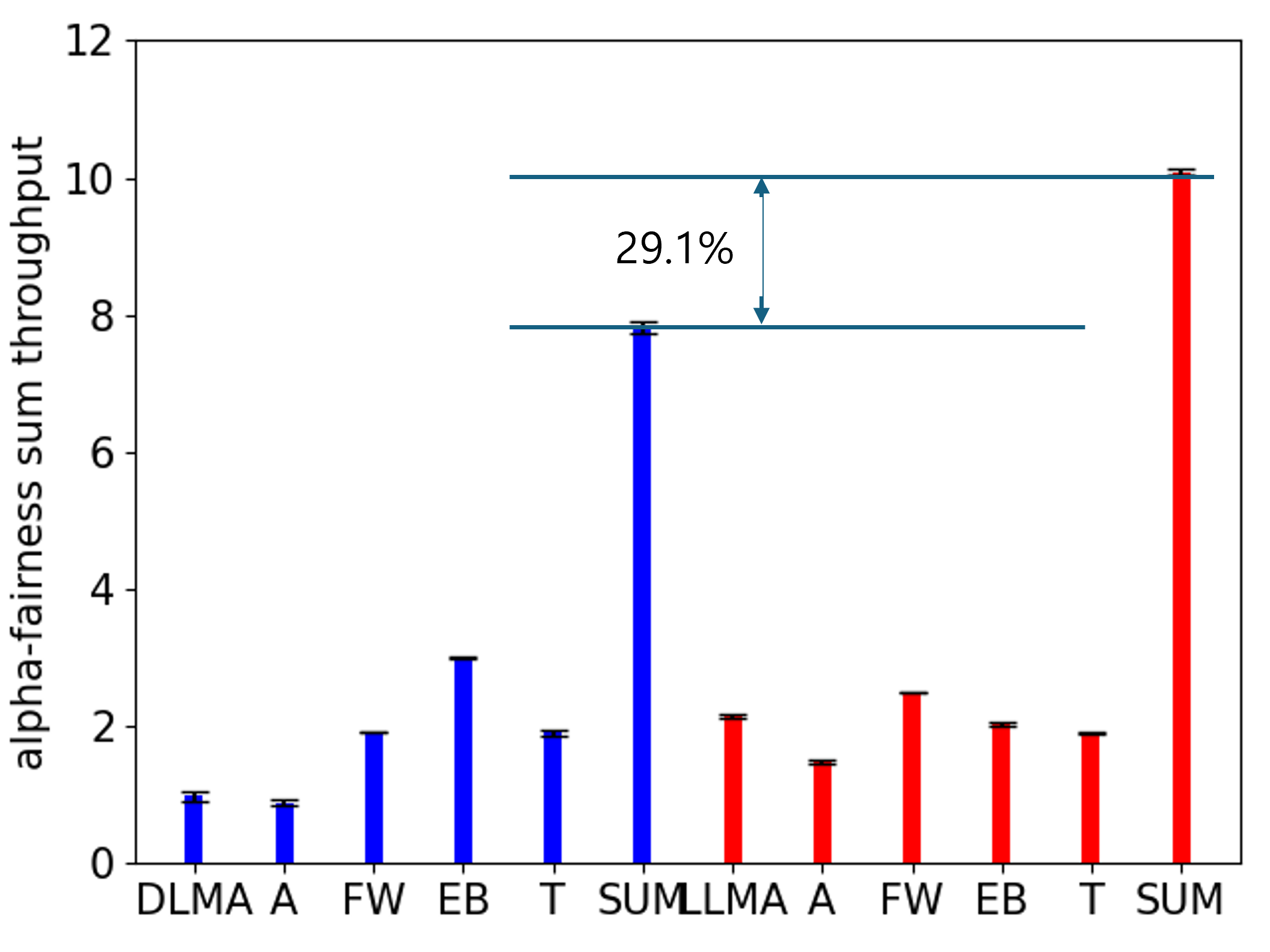}
         \caption{1H+1T+1A+1EB+1FW}
         \label{fig:1H+1T+1A+1EB+1FW}
     \end{subfigure}
     \hfill
     \begin{subfigure}[b]{0.48\columnwidth}
         \centering
         \includegraphics[width=\textwidth]{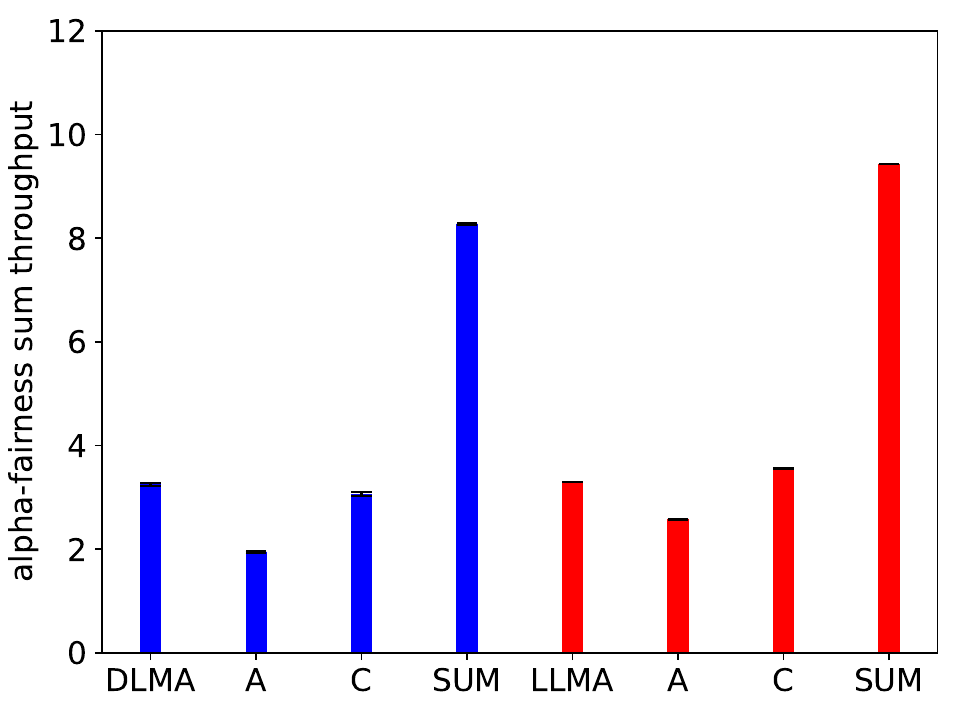}
         \caption{1H+1A+1C}
         \label{fig:1H+1A+1C}
     \end{subfigure}
     \hfill
     \begin{subfigure}[b]{0.48\columnwidth}
         \centering
         \includegraphics[width=\textwidth]{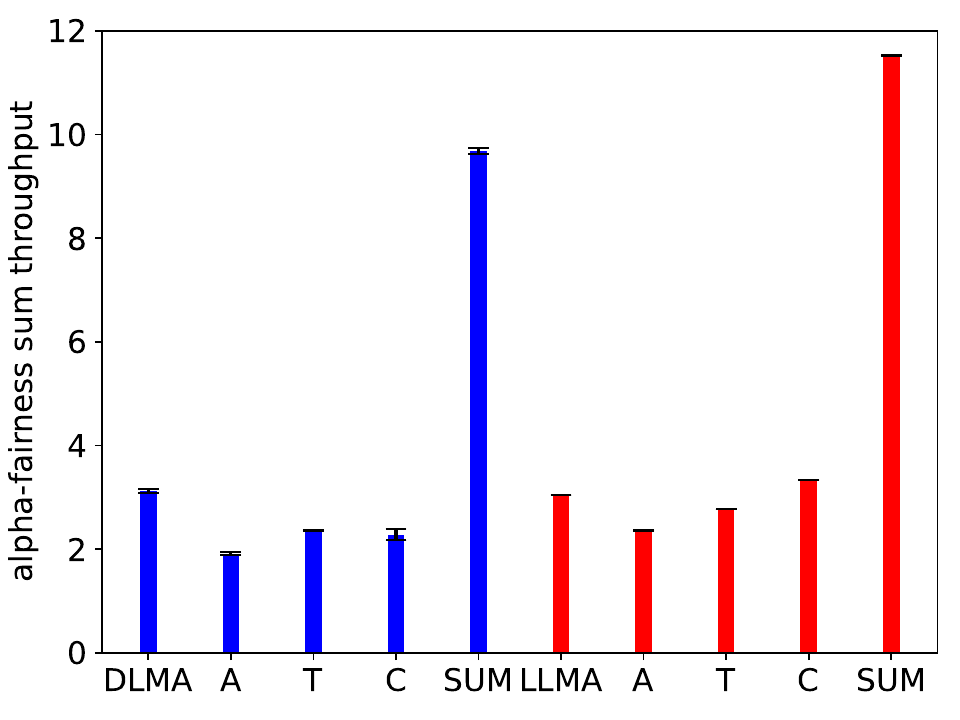}
         \caption{1H+1T+1A+1C}
         \label{fig:1H+1T+1A+1C}
     \end{subfigure}
    \setlength{\belowcaptionskip}{-20pt}
    \caption{$\alpha$-fairness throughput in a complex heterogeneous environment. C, EB, and FW represent CSMA node, EB-ALOHA node, and FW-ALOHA node, respectively.}
    \label{fig:Complex heterogeneous}
\end{figure}

We assess whether the LLMA node can function effectively in more complex scenarios.
Initially, we increase the number of ALOHA nodes to 20 to test the robustness of the LLMA node, setting the parameter $q=0.02$ for these ALOHA nodes. In this scenario of increased complexity, the LLMA node achieves near-optimal performance, evidenced by an RMSE value of 0.044, which is lower than the 0.052 observed for the DLMA node. 
Additionally, we introduce nodes using various other protocols, such as CSMA, EB-ALOHA, and FW-ALOHA. Since the optimal behavior of the AWARE node in coexistence with ALOHA, TDMA, CSMA, EB-ALOHA, and FW-ALOHA nodes has not been derived, we directly compare the performance of LLMA nodes against DLMA nodes by assessing the sum of $\alpha$-fairness throughput. The $\alpha$-fairness throughput of each node and the sum of $\alpha$-fairness throughput are presented in Fig.~\ref{fig:Complex heterogeneous}. To calculate $\alpha$-fairness throughput, we multiply the throughput by 100 to avoid negative values when applying the logarithm. Given our goal to maximize the sum of $\alpha$-fairness throughput, these figures demonstrate that the LLMA node outperforms the DLMA node by approximately 17.7\%, 29.1\%, 14.1\%, and 19.1\%, respectively. These results indicate that the LLMA protocol, designed through the CP-AgentNet, can effectively coexist in more complex heterogeneous environments.

\subsubsection{\textbf{Dynamic scenario}}
We evaluate the adaptability of the LLMA node in a dynamic setting to assess its responsiveness to environmental changes. This scenario starts with two slotted ALOHA nodes and one heteronode (2A+1H). At frame 2500, one ALOHA node exits the network (1A+1H), followed by the entry of two new ALOHA nodes at frame 5000 (3A+1H). The scenario progresses with the addition of one TDMA node at frame 7500 (1T+3A+1H), concluding at frame 10000. In this dynamic environment, as depicted in Fig.~\ref{fig:single_dynamic}, the operation of the LLMA node closely mirrors the ideal behavior. The RMSE value for the LLMA node is 0.044, which is approximately 5 times lower than 0.227 with the DLMA node, demonstrating that the LLMA, designed using the CP-AgentNet framework, effectively adapts to environmental changes. This significant difference arises because LLMA does not depend on the model architecture for different numbers of nodes. Unlike traditional DRL methods, the LLMA does not rely on a specific model architecture, thereby enhancing its scalability.

\begin{table}[t!]
  \centering
  \centering
  \caption{RMSE of all scenarios. T, A, and H represent TDMA, slotted ALOHA, and heteronode respectively.}
    \begin{tabular}{cccc}
        \toprule
          &Combination\hspace{2mm}     &DLMA\hspace{2mm}    &LLMA(ours)\hspace{2mm} \\
        \cmidrule(rl){1-4}
         %\multirow{3}{*}{\shortstack{Homogeneous LLMA}}
         %               &2H\hspace{2mm} &-\hspace{2mm} &0.0203\hspace{2mm} \\
         %               &4H\hspace{2mm} &-\hspace{2mm} &0.0148\hspace{2mm} \\
         %               &8H\hspace{2mm} &-\hspace{2mm} &0.0092\hspace{2mm} \\
        \cmidrule(rl){1-4} 
         \multirow{7}{*}{\shortstack{Single-heteronode}}
                        &1T+1H\hspace{2mm} &0.1409\hspace{2mm} &0.1908\hspace{2mm} \\
                        &1A+1H\hspace{2mm} &0.1175\hspace{2mm} &0.0768\hspace{2mm} \\
                        &2A+1H\hspace{2mm} &0.1192\hspace{2mm} &0.0491\hspace{2mm} \\
                        &3A+1H\hspace{2mm} &0.0931\hspace{2mm} &0.0468\hspace{2mm} \\
                        &4A+1H\hspace{2mm} &0.1044\hspace{2mm} &0.0433\hspace{2mm} \\
                        &1T+1A+1H\hspace{2mm} &0.0909\hspace{2mm} &0.0681\hspace{2mm} \\
                        &1T+3A+1H\hspace{2mm} &0.0544\hspace{2mm} &0.0403\hspace{2mm} \\
        \cmidrule(rl){1-4} 
         \multirow{2}{*}{\shortstack{Multi-heteronodes}}
                        &2A+2H\hspace{2mm} &0.0537\hspace{2mm} &0.0272\hspace{2mm} \\
                        &1T+2A+3H\hspace{2mm} &0.019\hspace{2mm} &0.0177\hspace{2mm} \\
        \cmidrule(rl){1-4} 
         \multirow{1}{*}{\shortstack{Massive nodes}}
                        &20A+1H\hspace{2mm} &0.0522\hspace{2mm} &0.044\hspace{2mm} \\
        \cmidrule(rl){1-4} 
         \multirow{1}{*}{\shortstack{Dynamic}}
                        &Dynamic\hspace{2mm} &0.227\hspace{2mm} &0.0476\hspace{2mm} \\
        \bottomrule
    \end{tabular}
  \label{tab:RMSE}
\end{table}

\subsubsection{\textbf{Ablation Study}}
\textbf{}
\\
\textbf{Impact of CP-Agent.} Although we have demonstrated the superior performance of the LLMA, it could be argued that this success stems from the pretrained knowledge of LLM. To demonstrate that the exceptional performance is specifically attributable to our CP-AgentNet, we conduct targeted assessments. Initially, we simulated the system using only the LLM under the same conditions as those used in CP-AgentNet. Subsequently, we evaluated the impact of each agent within the framework. Given the central role of the node agent in decision-making, we could not omit it from our simulations; therefore, we first ran simulations with the node agent paired with the strategy agent, and then with the observer agent, to isolate and assess their individual contributions. As shown in Table \ref{tab:CP impact}, in static tests, the RMSE values for both the LLMA node and the ALOHA node are similar between the ``No Agent" and ``Observer + Node" configurations, as well as between ``Strategy + Node" and ``Strategy + Observer + Node". This suggests that the strategy agent significantly influences the performance of both the LLMA and ALOHA. Conversely, the RMSE value for the TDMA node is notably affected by the observer agent, because the node agent avoids the time slots used by the TDMA node with the assistance of the observer agent. In dynamic scenarios, the observer agent impacts not only the behavior of the TDMA node but also that of the LLMA and ALOHA nodes, as it detects environmental changes and assists the node agent in quickly adapting to new conditions. These experiments have demonstrated the impact of each agent, with the superior performance originating from the CP-AgentNet framework. \\
%xyz: More experimental studies are needed to verify the model architecture design.
\textbf{Effectiveness of LLM ranker.} We evaluate the effectiveness of the LLM ranker. Specifically, during the offline stage, the strategy agent utilizes the LLM ranker to formulate transmission strategies. We conduct five simulations for each strategy, both with and without the LLM ranker, comparing the resultant RMSE values and their variances to assess which approach more closely approximates ideal behavior. In the online stage, we implement a transmission strategy that was formulated using the LLM ranker. Additionally, we explore scenarios where the node agent either uses or does not use the ranker to determine transmission probabilities. As shown in Table \ref{tab:llm_ranker}, both the average RMSE and standard deviation are lower when the strategy agent employs the LLM ranker, indicating enhanced stability and consistency. However, the impact of the LLM ranker is less critical for the node agent during the online stage. This reduced impact is because the node agent's decisions largely depend on the predefined strategy, which is relatively simpler than strategy formulation.

\begin{table}[t!]
\centering
  \caption{The impact of each agent.}
  \label{tab:CP impact}
    \begin{tabular}{ccccccc}
        \toprule
        \multirow{2}{*}{}\hspace{0mm} & \multicolumn{3}{c}{Static}\hspace{0mm} & \multicolumn{3}{c}{Dynamic}\\
        \cmidrule(rr){2-4}
        \cmidrule(rr){5-7}
          \hspace{0mm}      &LLMA\hspace{0mm} &ALOHA\hspace{0mm} &TDMA\hspace{0mm} &LLMA\hspace{0mm}   &ALOHA\hspace{0mm} &TDMA\\
        \cmidrule(rl){1-7} 
          1 \hspace{0mm} &0.138\hspace{0mm} &0.036\hspace{0mm} &0.058\hspace{0mm} &0.154\hspace{0mm} &0.04\hspace{0mm} &0.064\\
        \cmidrule(rl){1-7} 
          2 \hspace{0mm} &0.04\hspace{0mm} &0.015\hspace{0mm} &0.031\hspace{0mm} &0.089\hspace{0mm} &0.022\hspace{0mm} &0.033\\
        \cmidrule(rl){1-7} 
          3 \hspace{0mm} &0.155\hspace{0mm} &0.039\hspace{0mm} &0.011\hspace{0mm} &0.153\hspace{0mm} &0.041\hspace{0mm} &0.015\\
        \cmidrule(rl){1-7} 
          4 \hspace{0mm} &0.04\hspace{0mm} &0.015\hspace{0mm} &0.013\hspace{0mm} &0.048\hspace{0mm} &0.016\hspace{0mm} &0.01\\
        \bottomrule
    \end{tabular}
\end{table}

\begin{table}
\centering
  \caption{The impact of using an LLM ranker.}
  \label{tab:llm_ranker}
    \begin{tabular}{ccccc}
        \toprule
        \multirow{2}{*}{Phase}\hspace{3mm} & \multicolumn{2}{c}{Without Ranker}\hspace{3mm} & \multicolumn{2}{c}{With Ranker}\\
        \cmidrule(rr){2-3}
        \cmidrule(rr){4-5}
          \hspace{5mm}      &RMSE\hspace{3mm} & STD\hspace{3mm} & RMSE\hspace{3mm}   & STD\\
        \cmidrule(rl){1-5} 
          Offline learning \hspace{5mm} &0.0636\hspace{3mm} &0.036\hspace{3mm} &0.0404\hspace{3mm} &0.008\\
        \cmidrule(rl){1-5} 
          Online execution \hspace{5mm} &0.0362\hspace{3mm} &0.007\hspace{3mm} &0.0346\hspace{3mm} &0.007\\
        \bottomrule
    \end{tabular}
\end{table}

\subsubsection{\textbf{Explainability}}
\begin{figure}[t!]
     \centering
     \begin{subfigure}[b]{0.48\columnwidth}
        \centering
        \includegraphics[width=\textwidth]{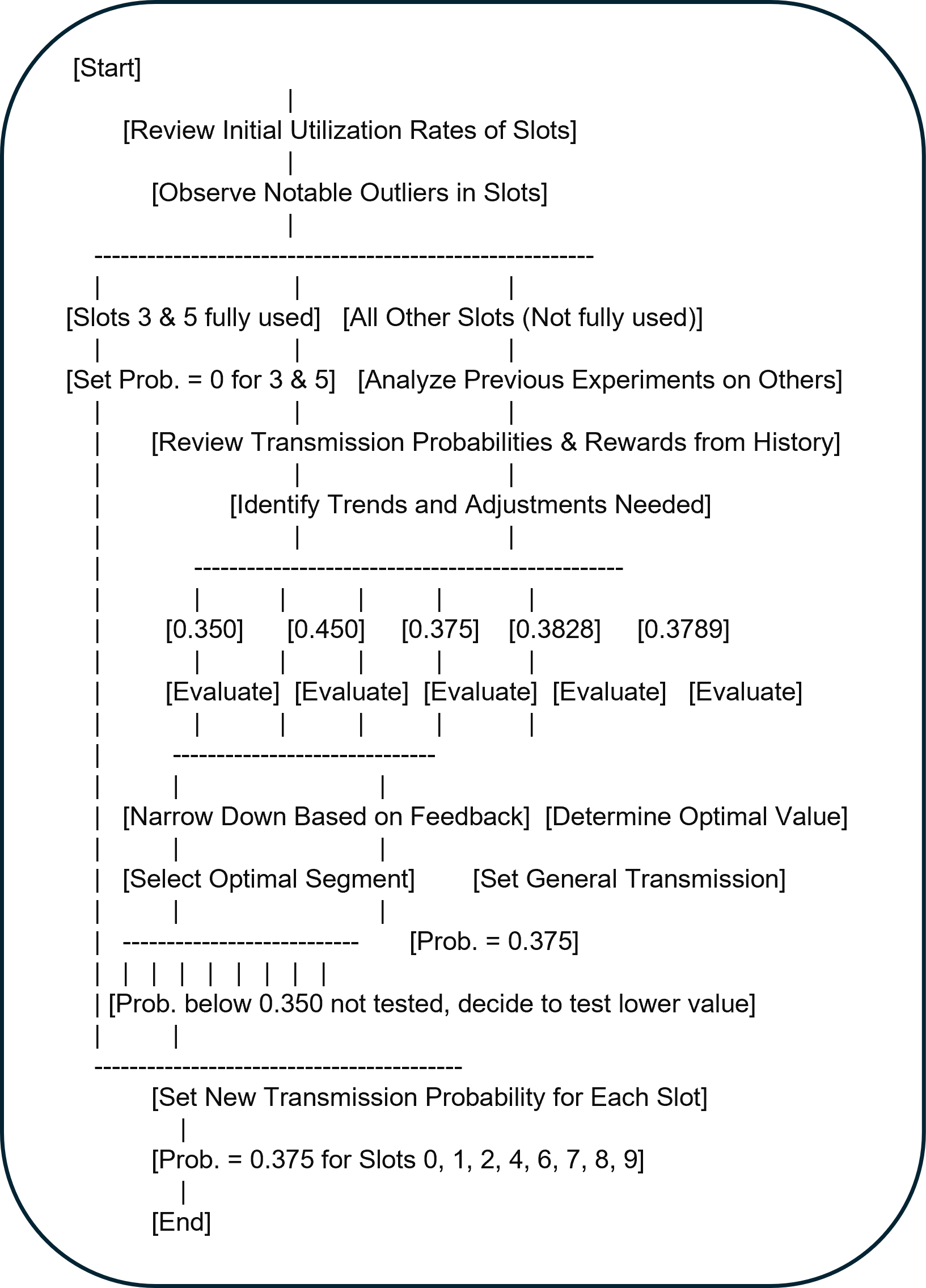}
        \subcaption{Decision tree on CP-Agent}
        \label{fig:decision_tree}
     \end{subfigure}
     \hfill
     \begin{subfigure}[b]{0.48\columnwidth}
        \centering
        \includegraphics[width=\textwidth]{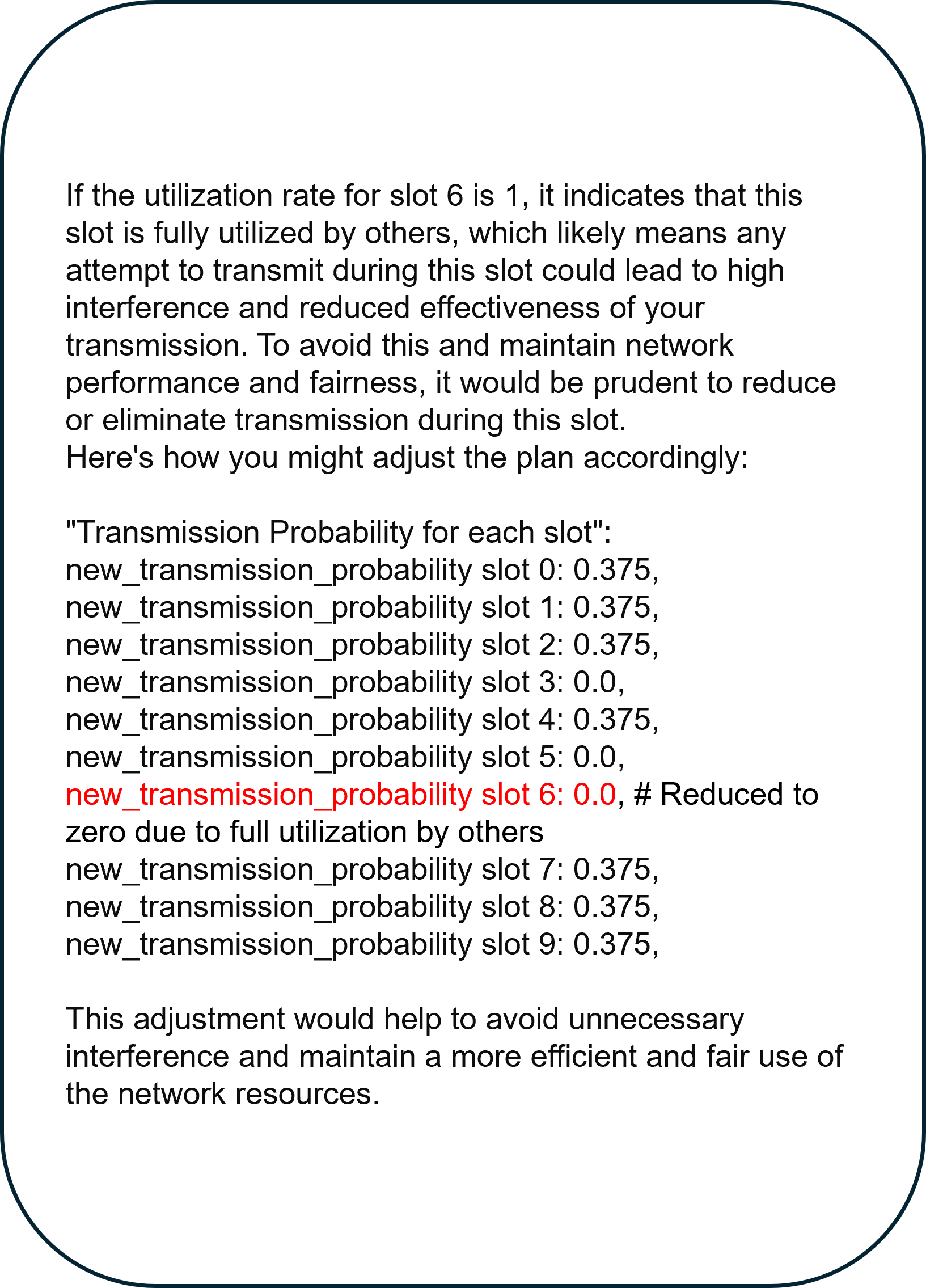}
        \subcaption{Ability to answer questions}
        \label{fig:whatif}
     \end{subfigure}
    \caption{Explainable CP-AgentNet}
\end{figure}

There is no established consensus on the definition of explainability in machine learning, nor on how it should be measured \cite{XAI}. Explainable AI (XAI) focuses on demystifying the decision-making processes of ML models—understanding how and why decisions are made\cite{XAI_web}. One of the most straightforward approaches to achieving explainability is to employ decision trees \cite{XAI}. In our framework, multi-agent interaction facilitates the generation of a decision tree, rendering the decision-making process transparent and interpretable. Fig.~\ref{fig:decision_tree} is generated by the LLM to illustrate how the action is devised. The observer agent monitors the environment and informs the node agent that slot 3 and 5 are fully utilized. Consequently, the node agent avoids these slots and determines the action for each slot. This process is clearly depicted in the decision tree, enabling users to easily understand how the decisions are made. Moreover, the LLM facilitates the \textit{ability to question}, enhancing comprehension of the decisions for laypeople and enabling them to explore `what-if' scenarios— a capability not available with traditional DRL methods. In our scenario, the LLM can respond to queries like `What if the utilization rate of slot 6 is 1?' without the need for additional training. The response is illustrated in Fig.~\ref{fig:whatif}, demonstrating the capability to facilitate relevant questioning.

\subsubsection{\textbf{Computational Overhead}} 
We evaluate the computational overhead for DLMA and LLMA nodes. Due to LLMA's reliance on the OpenAI API, direct hardware comparisons with DLMA are impractical. Therefore, DLMA's performance is assessed across various GPUs (TPU v2, A100, L4, and A6000). LLMA consistently demonstrates faster decision-making compared to DLMA. Specifically, DLMA required approximately 841 seconds on the fastest GPU (A100) to process decisions for 1000 frames (10 seconds), whereas LLMA achieved the same in approximately 19 seconds for a query period of one second. While neither DLMA nor LLMA currently meets real-time processing requirements, LLMA's performance significantly outpaces DLMA, highlighting its strong potential for future real-time applications.

\subsection{CPTCP results}

\begin{table}
\centering
  \caption{Total throughput (Kbps) with homogeneous TCP}
  \label{tab:homoTCP}
    \begin{tabular}{ccccccc}
        \toprule
          number of flows \hspace{1mm} &Reno\hspace{1mm} & Vegas\hspace{1mm} & CPTCP\hspace{1mm}   & LLM (no CP-Agent)\\
        \cmidrule(rl){1-5} 
          2 \hspace{1mm} &721.5\hspace{1mm} &765.2\hspace{1mm} &813.6\hspace{1mm} &543.2\\
        \cmidrule(rl){1-5} 
          3 \hspace{1mm} &840.9\hspace{1mm} &966.9\hspace{1mm} &868.4\hspace{1mm} &679.1\\
        \cmidrule(rl){1-5} 
          4 \hspace{1mm} &868.0\hspace{1mm} &964.1\hspace{1mm} &890.1\hspace{1mm} &584.0\\
        \bottomrule
    \end{tabular}
\end{table}

\subsubsection{\textbf{Homogeneous scenario}}
We evaluate multiple CPTCP flows in configurations of 2, 3, and 4 in a homogeneous environment, comparing the total throughput results with TCP Reno and TCP Vegas. As shown in Table \ref{tab:homoTCP}, although CPTCP does not overwhelmingly outperform Reno and Vegas, these results demonstrate that comparable TCP performance can be achieved through CP-AgentNet. Importantly, when the LLM is used without CP-Agent, performance consistently ranks as the lowest, regardless of the number of flows. These results indicate that the performance of the CPTCP does not stem solely from the pretrained LLM. 

\subsubsection{\textbf{Heterogeneous scenario}}
To demonstrate that CPTCP can coexist with different types of TCP algorithms, we assess its performance in a heterogeneous TCP environment. As highlighted in the gray area of Table \ref{tab:heteroTCP}, when loss-based TCP (Reno) and delay-based TCP (Vegas) coexist, a fairness issue arises, resulting in a low Jain's fairness index of 0.796. However, when CPTCP coexists with either Reno or Vegas, it does not interfere with the other TCP algorithm, achieving Jain's fairness indices of 0.994 and 0.983. This indicates that CPTCP can coexist effectively with both types of TCP algorithms.

\begin{table}[t!]
\centering
\setlength{\belowcaptionskip}{10pt}
\caption{Throughput and fairness with heterogeneous TCP}
\label{tab:heteroTCP}
\renewcommand{\arraystretch}{1.2} % Adjust row height
\setlength{\tabcolsep}{5pt} % Adjust column separation
\begin{tabular}{ccccc}
\toprule
\multicolumn{2}{c}{TCP algorithm} & \multicolumn{2}{c}{Throughput (Kbps)} & \multirow{2}{*}[-0.3em]{Jain's fairness index} \\ \cmidrule(rr){1-2} \cmidrule(rr){3-4} 
flow1 & flow2 & flow1 & flow2 &  \\ \midrule
Reno & Reno & 377.3 & 344.2 & 0.998 \\ \midrule
Vegas & Vegas & 382.8 & 382.4 & 1.000 \\ \midrule
CPTCP & CPTCP & 398.6 & 415.0 & 1.000 \\ \midrule
Reno & Vegas & \cellcolor{gray}589.7 & \cellcolor{gray}193.6 & \cellcolor{gray}0.796 \\ \midrule
Reno & CPTCP & \cellcolor{gray}376.2 & \cellcolor{gray}442.4 & \cellcolor{gray}0.994 \\ \midrule
Vegas & CPTCP & \cellcolor{gray}440.1 & \cellcolor{gray}336.6 & \cellcolor{gray}0.983 \\ \bottomrule
\end{tabular}
\end{table}

%To demonstrate that CPTCP can coexist with different types of TCP flows, we evaluated the performance of all combinations of Reno, Vegas, and CPTCP with two flows. While each flow exhibits similar throughput in homogeneous TCP environments, significant interference is observed when Reno and Vegas flows coexist, as detailed in Table [reference needed]. Specifically, the throughput of Vegas drops by approximately 49\%, resulting in the lowest sum of $\alpha$-fairness among all combinations. In contrast, Vegas's throughput does not decline when it coexists with CPTCP. Moreover, CPTCP manages to coexist with both Vegas and Reno without being adversely affected by Reno. In summary, the sum of $\alpha$-fairness is superior when CPTCP coexists with other TCP flows compared to when Reno coexists with Vegas.

\section{Related Works}\label{Related_works}

\subsection{LLM-empowered Agents}
%With the evolution of autonomous agents empowerd by LLMs, complex tasks such as game playing, software developments, and reasoning are addressed. 
Generative Agents \cite{generative_agent} elicit emerging behaviors and collaboration within assemblies of agents as societal groups. GITM \cite{ghost} focuses on developing Generally Capable Agents (GCAs) using LLMs within the Minecraft video game. %In this initiative, the agent breaks down high-level plans into sub-goals. It then sequentially executes actions to address these sub-goals, ultimately completing the overall task. 
Autonomous software development has improved through the collaborative use of multiple LLM agents\cite{self-collaboration, agentverse, metagpt, unleashing, qian2023communicative, huang2023agentcoder,li2023camel}. These systems utilize various LLM agents, assigning specific tasks to each to reduce complexity and facilitate collaboration in accomplishing these tasks.  Notably, MetaGPT \cite{metagpt} and AgentVerse \cite{agentverse} simulate a human-like software development process, enabling efficient role-playing for each agent. Additionally, improvements in mathematical accuracy using LLM agents have been explored \cite{improving, dylan}. 
%Specifically, these agents integrate candidate outputs from peer agents with the initial problem statement, allowing them to review and refine their answers \cite{improving}. DyLAN \cite{dylan} models agents as neural networks where multiple iterations of agent interactions are depicted as layers within a network, and communication between agents is represented by neural connections. %By dynamically adjusting both the quantity and configuration of these layers, DyLAN effectively addresses not only mathematical challenges but also a variety of other complex tasks. 
While these LLM-agents are typically built on top of LLMs accessed via the internet, personal LLM-agents\cite{personalllm} can also be deployed on-device using LLM\cite{llmacd}. CP-AgentNet operates under the assumption that personal LLM-agents can be utilized, although in practice, we actually employ GPT-4-Turbo via the internet.

\subsection{LLMs with Reinforcement Learning}
Recent research has explored various ways of integrating reinforcement learning with LLMs\cite{when2ask,LLARP,incontext,sun2023offline, peters2007reinforcement, hu2023aligning, song2023self, kwon2023reward}. When2Ask \cite{when2ask} method involves using RL to control the frequency of queries to the LLM, optimizing for cost-effective interaction between the agent and the LLM. 
%In this approach, RL serves as a mediator block, determining when to consult the LLM based on changes in observations, thereby reducing both communication time and monetary costs. 
LLARP \cite{LLARP} employs a pre-trained, frozen LLM as the core neural network within a DRL framework. This setup enhances the agent's generalization capabilities for embodied tasks through the use of LLM. ICPI \cite{incontext} iteratively updates the contents of the prompt, which serves as the foundation for its policy, through trial-and-error interactions within an RL environment. %In this environment, ICPI chooses actions that maximize the estimated Q-value from the current state. 
This approach does not employ gradient-based methods; instead, the focal point of learning is the prompt content itself. 
%xyz: Is our model an "RL" model? Does our model satisfy the definition of an RL model? 
%dc: RL is the science of decision making. It is about learning the optimal behavior in an environment to obtain maximum reward. This optimal behavior is learned through interactions with the environment and observations of how it responds. From this point of view, our work satisfy the RL framewwork although it is exactly matched to the RL. (discuss)
%Our framework employs an LLM-driven RL strategy, utilizing LLMs as agents within an RL framework, and presents significant differences from existing approaches. While it similarly eschews the use of model parameters and gradient methods, our method fundamentally diverges by shifting the locus of learning from the prompt to the outputs of the LLMs, which are refined through a process of self-examination.

\subsection{LLMs for communication network} 
Following the trend of leveraging LLMs for specialized disciplines, there is a growing movement to adopt LLMs in the field of communication networks \cite{Bert-QA, LLM_standard, LLM_telecom, LLM_small, teleqna, TKG, telecom1, telecom2}. The work in \cite{LLM_telecom} fine-tunes LLMs with telecommunication-specific language to identify working groups within the 3rd Generation Partnership Project (3GPP). \cite{LLM_standard, Bert-QA, LLM_small} utilize LLMs as question answering (QA) assistants to interpret 3GPP standards. %Conversely, \cite{teleqna} introduces a dataset to assess the telecommunication knowledge of LLMs.
%while \cite{TKG} explores data governance and creates a domain-specific corpus to refine LLMs in the telecommunications field. 
Collectively, these approaches primarily focus on classification of working groups or QA tasks, indicating that the use cases of LLMs in the communication domain remain quite limited. 
%xyz: Clarify what's unique about this. No one studied this doesn't mean this is significant.
%dc: modified as below. Mentioned very briefly here because it is mentioned in the introducion part.
\section{Conclusion}\label{Conclusion}
We introduced CP-AgentNet, the first framework using generative agents to autonomously design communication network protocols, enhancing explainability. Using CP-AgentNet, we designed the LLMA and CPTCP for heterogeneous environments. Our simulations show that nodes using LLMA or CPTCP coexist efficiently with nodes using different protocols. This work sets the stage for broader applications of generative agents in protocol design and solving complex network problems.

%\bibliography{main.bib}
\bibliographystyle{ieeetr}

\end{document}